\documentclass[twocolumn,superscriptaddress,preprintnumbers,amsmath,amssymb,prl,longbibliography]{revtex4-1}
\usepackage{graphicx}
\usepackage{dcolumn,xcolor}
\usepackage{amsmath} 
\usepackage{amssymb}
\usepackage{amsfonts}
\usepackage{bm}
\usepackage[latin1]{inputenc}
\usepackage[dvips]{epsfig}
\usepackage{hyperref}
\usepackage{color}
 \usepackage[normalem]{ulem}

\definecolor{orange}{rgb}{1,0.6,0}

\begin{document}

\title{Tuning the shear-thickening of suspensions  through surface roughness and physico-chemical interactions}

 \author{Philippe Bourrianne}
    \affiliation{Hatsopoulos Microfluids Laboratory, Department of Mechanical Engineering, MIT, 77 Massachusetts Avenue, Cambridge, MA 02139, USA}
 \author{Vincent Niggel}
    \affiliation{Hatsopoulos Microfluids Laboratory, Department of Mechanical Engineering, MIT, 77 Massachusetts Avenue, Cambridge, MA 02139, USA}
 \author{Gatien Polly}
    \affiliation{Hatsopoulos Microfluids Laboratory, Department of Mechanical Engineering, MIT, 77 Massachusetts Avenue, Cambridge, MA 02139, USA}
    \affiliation{MultiScale Material Science for Energy and Environment, UMI 3466, CNRS-MIT, 77 Massachusetts Avenue, Cambridge, MA 02139, USA}
 \author{Thibaut Divoux}
    \affiliation{MultiScale Material Science for Energy and Environment, UMI 3466, CNRS-MIT, 77 Massachusetts Avenue, Cambridge, MA 02139, USA}
    \affiliation{ENSL, CNRS, Laboratoire de physique, F-69342 Lyon, France}
\author{Gareth H. McKinley}
    \affiliation{Hatsopoulos Microfluids Laboratory, Department of Mechanical Engineering, MIT, 77 Massachusetts Avenue, Cambridge, MA 02139, USA}

\date{\today}

\begin{abstract}
Shear thickening denotes the reversible increase in viscosity of a suspension of rigid particles under external shear. This ubiquitous phenomenon has been documented in a broad variety of multiphase particulate systems, while its microscopic origin has been successively attributed to hydrodynamic interactions and frictional contact between particles. The relative contribution of these two phenomena to the magnitude of shear thickening is still highly debated and we report here a discriminating experimental study using a model shear-thickening suspension that allows us to tune independently both the surface chemistry and the surface roughness of the particles. We show here that both properties matter when it comes to continuous shear thickening (CST) and that the presence of hydrogen bonds between the particles is essential to achieve discontinuous shear thickening (DST) by enhancing solid friction between closely contacting particles. Moreover, a simple argument allows us to predict the onset of CST, which for these very rough particles occurs at a critical volume fraction much lower than that previously reported in the literature. Finally, we demonstrate how mixtures of particles with opposing surface chemistry make it possible to finely tune the shear-thickening response of the suspension at a fixed volume fraction, paving the way for a fine control of the shear-thickening transition in engineering applications. 
\end{abstract}

\maketitle

Suspensions composed of rigid particles dispersed in a liquid matrix are ubiquitous in daily life, from cosmetics and consumer products (e.g., tooth paste) to foodstuffs and engineering materials such as fresh cement paste. Their flow properties display a broad variety of rheological behavior depending on the volume fraction $\phi$ and the nature of the physico-chemical interactions between particles \cite{Stickel:2005,Mewis:2011}. 
In the semi-dilute regime, suspensions show a shear-thinning response, i.e., their shear viscosity $\eta$ decreases for increasing applied shear rate $\dot \gamma$. Such behavior, which results from long-range hydrodynamic interactions and possible shear-induced structure formation \cite{Cohen:2004,Cheng:2012,Vazquez:2016,Varga:2019}, is amplified in the presence of attractive interactions between particles. At moderately high volume fraction, these attractive interactions eventually confer a yield stress to the suspension, i.e., the material becomes paste-like with a solid-like elasto-plastic behavior at rest \cite{Bonn:2017}.

So-called ``dense suspensions" (i.e., suspensions with high volume fraction) exhibit shear thinning under weak shearing conditions followed by a shear-thickening transition for large enough shear, defined by a reversible increase of viscosity with the shear rate, or shear stress~\cite{Barnes:1989b, Wagner:2009, Denn:2018}. The magnitude of the shear thickening is enhanced with increases of the volume fraction. At intermediate volume fractions, continuous shear thickening (CST) ranges from a mild increase in viscosity that was initially explained by lubricated hard-sphere interactions~\cite{Bossis:1989, Jamali:2019} to a more pronounced effect when particle interactions involve frictional contacts~\cite{Seto:2013}. Finally, for very large volume fractions, closer to the jamming volume fraction ${\phi}_{\rm{J}}$~\cite{Liu:2010}, the viscosity may jump by orders of magnitude at a constant shear rate \cite{Wyart:2014,Mari:2014}. Such a discontinuous shear thickening (DST) response was first associated with the formation of ``hydroclusters" \cite{Brady:1985, Bender:1996, Melrose:2004, Egres:2005, Fernandez:2013}, before being attributed primarily to frictional interactions  \cite{Boyer:2011,Seto:2013,Wyart:2014,Pan:2015,Comtet:2017,Clavaud:2017,Hsu:2018,Morris:2018} or even shear-jamming regimes \cite{Peters:2016}. The nature of local inter-particle interactions, mediated by the suspending solvent, affects this picture. Indeed, the shear-thickening transition is enhanced when particles show moderate attraction \cite{Park:2019}, and suppressed in the absence of repulsive forces \cite{Clavaud:2017} or in presence of strong short-range interparticle attractions leading to the presence of a yield stress \cite{Gopalakrishnan:2004,Brown:2010}. More recently, hydrogen bonding interactions have been shown to enhance DST by favoring weak and reversible adhesion between particles \cite{Comtet:2017,James:2018,James:2019b,Son:2019}.

\begin{figure*}[t!]
\centering
\includegraphics[width=.8\linewidth]{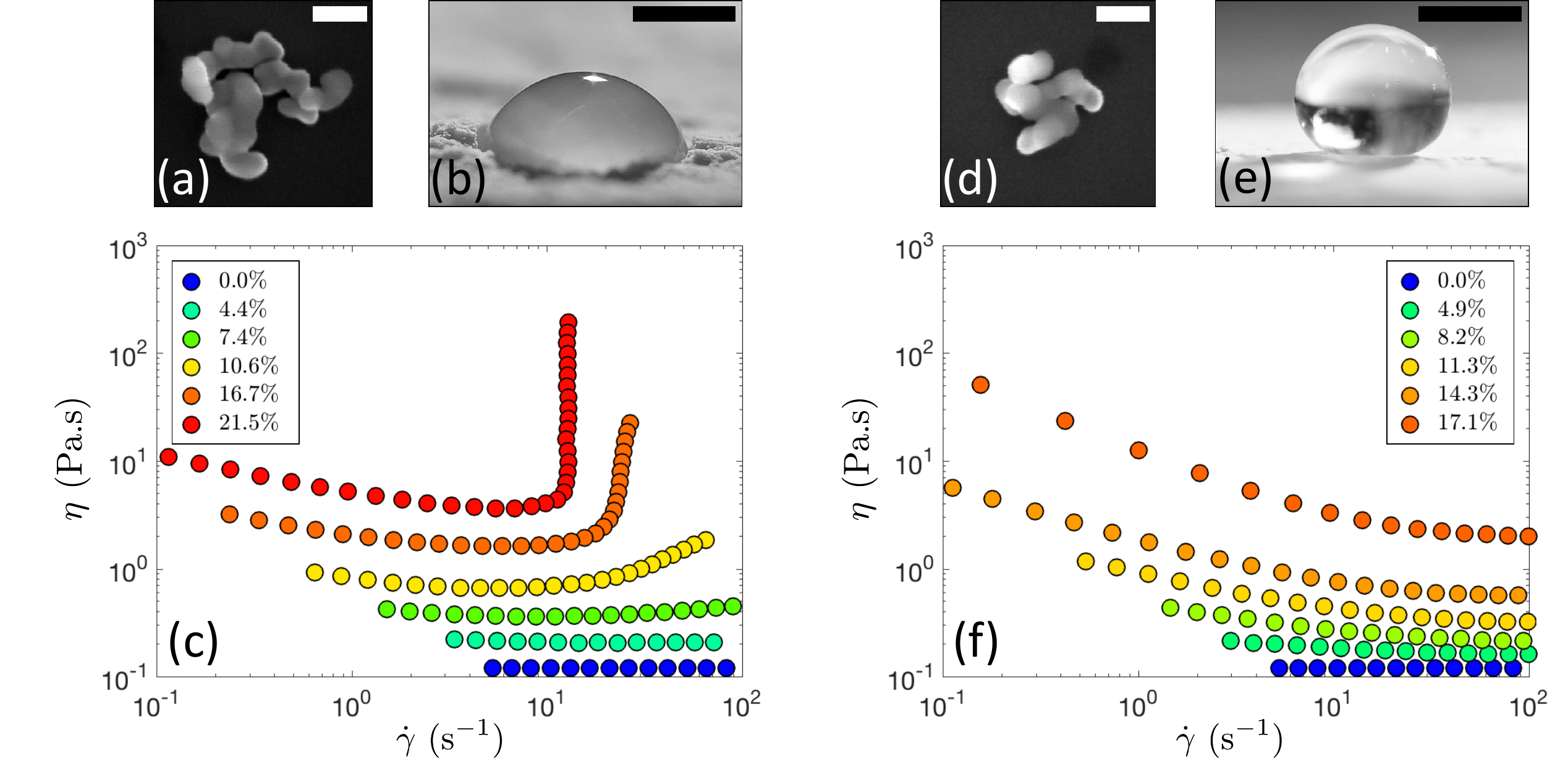}
\caption{(color online) (a) Scanning Electron Microscopy (SEM) images of a hydrophilic fumed silica particle. The scale bar represents 100 nm. (b) Hydrophilic fumed silica particles coated on a glass slide are wetted by a drop of water due to their hydrophilic nature. The scale bar is 1 mm. (c) Flow curves $\eta(\dot \gamma)$ measured for suspensions of hydrophilic fumed silica particles at various volume fractions $0 \le \phi \le 21.5{\%}$. The suspensions of hydrophilic particles show successively CST, and then DST, for increasing $\phi$. (d) SEM picture of a hydrophobic fumed silica particle. The scale bar represents 100 nm. (e) A water drop is repelled by a coating made by depositing a layer of hydrophobic fumed silica particles on a glass slide. The superhydrophobic character of the coating highlights the hydrophobic chemistry of the particles. The scale bar represents 1 mm. (f) Viscosity $\eta$ of a suspension of hydrophobic fumed silica as a function of shear rate $\dot \gamma$ at various volume fractions $\phi$. The suspensions of hydrophobic particles display a shear-thinning response for all the volume fractions tested.
\label{fig1}}
\end{figure*} 

To date DST has, in general, been associated with dense suspensions, i.e., it occurs for $\phi$ relatively close to the jamming fraction $\phi_{\rm J}$ \cite{Lootens:2003,Brown:2009,Morris:2018}. In this framework, the current mechanistic interpretation of DST relies on the divergence of the suspension viscosity in the vicinity of $\phi_{\rm J}$~\cite{Wyart:2014, guazzelli2018rheology}. While repulsive forces between particles dominate at low shear-rate, particles are brought in contact under shear in a transition to a frictional regime~\cite{Wyart:2014,morris2020shear}. Despite an emerging consensus on that frictional scenario, recent studies have attributed the origin of the frictional interactions to, respectively, solid friction~\cite{Clavaud:2017}, weak chemical attractive interactions~\cite{Comtet:2017, James:2018} or hydrodynamic forces~\cite{Jamali:2019}. The relative contributions of the solid friction and the chemical interactions at the molecular scale to the magnitude and dynamics of the DST remain to be quantified, but are both embodied in the magnitude of the friction coefficient between particles. Here we report an experimental study on suspensions of colloidal fumed silica particles suspended in a viscous oligomeric polar solvent. We show that these suspensions, which do not exhibit any ageing, drying, migration or settling (in sharp contrast to cornstarch or other non-Brownian suspensions~\cite{kusina2021aging}) can display both CST and DST transitions at much lower volume fractions than the values traditionally reported in the literature. Furthermore, these particles allow us to tune independently their surface chemistry and their effective surface roughness, which makes it possible to identify  and discriminate experimentally the relative contributions of both friction and reversible chemical adhesion to the shear-thickening transition. We demonstrate that reducing frictional interactions fully suppresses the shear-thickening transition. Finally, by mixing hydrophilic and hydrophobic particles in varying fractions, we study the competition between initiation and inhibition of the discontinuous shear-thickening transition, which allows us to smoothly tune the onset and extent of DST.

To study the shear thickening transition we utilize four distinct fumed silica systems that enable us to vary both the  surface roughness and the hydrophilicity of the particles while holding the average particle size constant at $D\simeq 300$~nm. These textured fumed silica particles are composed of nodules of size $R_{\rm u}$ that are fused together permanently during the flame synthesis \cite{Barthel:1998}. The sintered aggregates confer nanometric textures and high specific surface areas (measured in $\rm{m^2/g}$) to the particles [see Fig.~\ref{fig1}(a) and Fig.~\ref{figS1}]. We consider here two classes of fumed silica particles, which differ by the size $R_{\rm{u}}$ of the primary unit and consequently by their nanometric surface roughness. A smaller primary unit ($R_{\rm u}$ = 10 nm, later denoted compactly as ``rougher" fumed silica) induces  a more convoluted texture (i.e., higher specific area) than larger nodules ($R_{\rm u}$ = 25 nm, ``less rough" fumed silica). Both of these types of particles are naturally strongly hydrophilic due to both the presence of surface hydroxyl groups and to their intrinsic submicronic roughness as seen in the picture in Fig.~\ref{fig1}(b). When dispersed in polypropylene glycol (PPG, $\eta _0 = 120$ mPa.s, ${M}_w = 725$ g/mol), i.e., a polar solvent, these fumed silica particles display a stabilizing solvation layer that prevents irreversible aggregation \cite{Raghavan:1997} indicating the presence of a short-range repulsive force, in addition to thermal fluctuations, which have also been suggested as a possible repulsive force in Brownian suspensions~\cite{mari2015discontinuous, pednekar2017simulation}. The resulting suspensions  thus show excellent stability over time with no thixotropy or ageing (see Fig.~\ref{figS2})  as well as exhibiting optical transparency. We have prepared suspensions at various volume fractions $\phi$ by varying the mass ratio of particles dispersed. The true volume fraction of the particulate phase is carefully determined by performing precise density measurements of the suspension (see  details in section~\ref{figS3}). Finally, the steady rheological response of the suspensions is measured by a slowly decreasing ramp of shear stress imposed in a cone-plate geometry connected to a stress-controlled rheometer (see section~\ref{figS4}). 

\begin{figure*}[t!]
\centering
\includegraphics[width=0.9\linewidth]{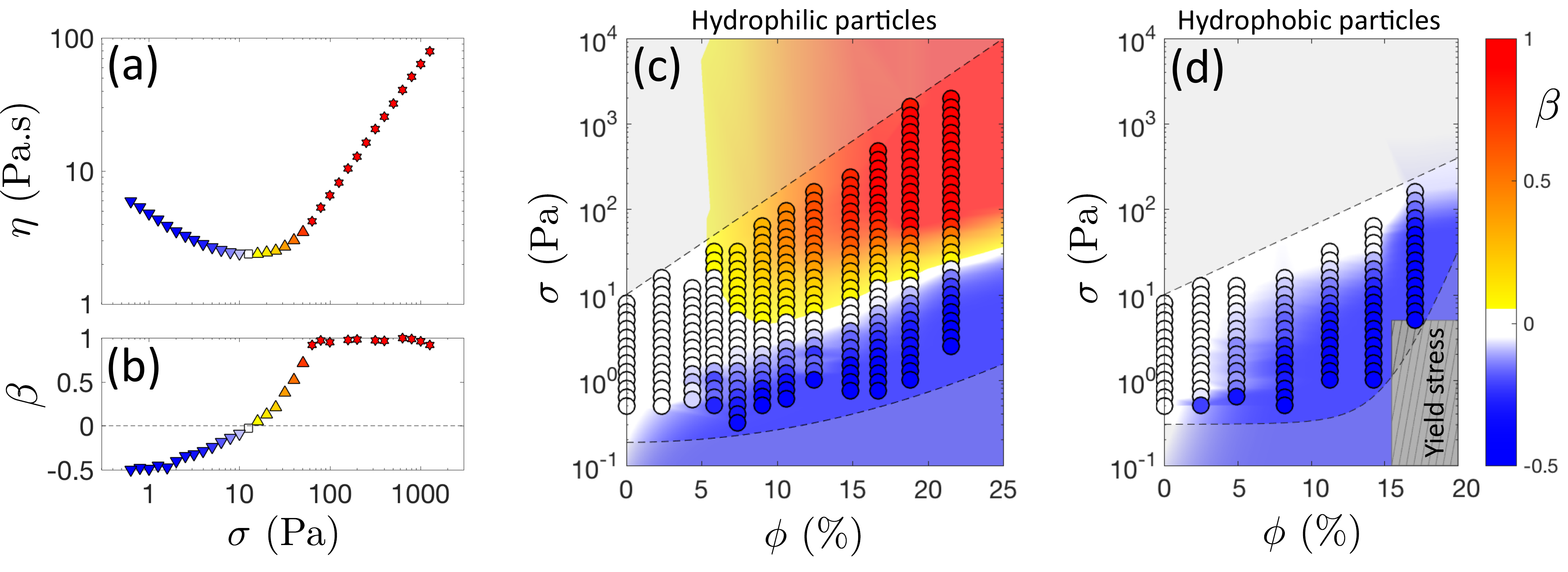}
\caption{(color online) (a)  Steady-state viscosity $\eta$ and (b) slope parameter $\beta=\dot \gamma ({\rm d}\eta/{\rm d}\sigma)$ vs. shear stress $\sigma$ for a suspension of hydrophilic fumed silica particles ($\phi = 18.9$\%, $R_{\rm u}=25$~nm). Different regimes are defined with respect to $\beta$: shear thinning ($\beta < 0$, \textcolor{blue}{$\blacktriangledown$}), Newtonian ($\beta= 0$, $\square$), CST ($0 \leq \beta < 1$, \textcolor{orange}{$\blacktriangle$}) and DST  ($\beta \to 1$, \textcolor{red}{$\star$}). 
(c)~Phase diagram ($\sigma$, $\phi$) for hydrophilic fumed silica particles. The hollow black circles represent individual steady-state rheological measurements and colors encode the values of $\beta$. We used the same color code as in (a) and (b). (d)~Phase diagram ($\sigma$, $\phi$) for hydrophobic fumed silica particles. Same color code as in (a)-(c). Suspensions of hydrophobic fumed silica do not exhibit any shear-thickening transition. The grey hatched rectangle at low stress and high volume fraction denotes the region influenced by the yield stress. In both (c) and (d), the two areas shaded in light grey transparency at low and high shear stress correspond to regions that cannot be accessed experimentally (see section~\ref{figS6}). 
\label{fig2}}
\end{figure*} 

The rheometric response of the less rough ($R_{\rm u}$ = 25 nm) hydrophilic fumed-silica suspensions is presented in Fig.~\ref{fig1}(c) for volume fractions ranging between $\phi=0$\% and 21.5\%. Increasing the mass of dispersed particles leads to a departure from the Newtonian response of the pure solvent ($\eta_0 = 120$~mPa.s). At low volume fractions, the suspension shows a weakly shear-thinning behavior characterized by a slow decrease of the viscosity $\eta$ for increasing shear rates $\dot \gamma$. At intermediate volume fractions ($\phi \simeq 7.4$\%), the shear-thinning behavior persists at low shear rates ($\dot \gamma < 20$~s$^{-1}$), whereas the viscosity begins to slowly increase at high shear rates, which is characteristic of CST \cite{Brady:1985,Brown:2009}. This trend is amplified for increasing volume fractions of fumed silica, up to $\phi = 16.7$\% for which the viscosity jumps by an order of magnitude over a narrow range of shear rates. Finally, for even larger volume fractions ($\phi = 21.5$\%), an abrupt shear-thickening transition occurs at an almost constant critical shear rate of about 15~s$^{-1}$, which corresponds to DST \cite{Barnes:1989b,Brown:2012,Seto:2013}. In this suspension, the CST and DST transitions thus occur respectively at $\phi_{\rm{CST}} \simeq 5{\%}$ and $\phi_{\rm{DST}} \simeq 18{\%}$. Strikingly, these volume fractions are significantly lower than that reported in the literature, where shear thickening is usually observed for $\phi \gtrsim 50$\% in the vicinity of the jamming point $\phi_{\rm J}$, in dense suspensions of both Brownian and non-Brownian particles with more regular shapes \cite{Mewis:2011,Pan:2015,Royer:2016,Clavaud:2017} although the precise value for the transition is affected systematically by the particle geometry \cite{James:2019}. This result suggests the presence of strong dynamic interactions between fumed silica particles at packing fractions much lower than the jamming fraction $\phi_{\rm J}$, which must arise from the complex structural geometry of our particles. In a similar way, the fractal-like structure of carbon black particles leads to suspensions that exhibit weak shear thickening \cite{Osuji:2008} or the appearance of a yield stress behavior at correspondingly low volume fraction \cite{Richards:2017}. However, it is important to note that discontinuous shear thickening is not reported in carbon black suspensions. One important distinction is the hydrophobic nature of carbon black which typically necessitates the use of apolar organic solvents to generate stable dense suspensions\cite{Waarden:1950}.

In order to directly probe the effect of the particle surface chemistry, we repeat the same rheological measurements on suspensions of \textit{hydrophobic} fumed silica particles. These particles are coated with a nanometric layer of silanes. They display identical geometrical features ($R_{\rm u}$, $N_{\rm u}$ and $D$) as the hydrophilic particles [see Fig.~\ref{fig1}(d) and Fig.~\ref{figS1}]  and differ only in their surface chemistry. As we show in Fig.~\ref{fig1}(e), a surface coating of these particles on a glass slide results in a superhydrophobic surface with contact angle $\theta \ge 150^{\circ}$. At low volume fraction, those hydrophobic particles still disperse in polypropylene glycol. As observed in Fig.~\ref{fig1}(f), suspensions of hydrophobic fumed silica exhibit a strikingly different response from that observed with hydrophilic particles at the same volume fraction [Fig.~\ref{fig1}(c)]. These hydrophobic particulate suspensions display a purely shear-thinning response over the entire range of volume fractions explored. The shear-thinning effect becomes more pronounced for increasing volume fractions, in part due to the gradual appearance of a dynamic yield stress at large volume fractions ($\phi \ge 17.1$\%), whose quantification is achieved with additional oscillatory shear protocols as described in details in section~\ref{figS5}. However, none of the suspensions of hydrophobic particles investigated show any shear-thickening behavior at high imposed shear rates. This result shows that modifying the surface chemistry of the fumed silica particles from hydrophilic to hydrophobic fully suppresses the shear-thickening transition over the entire range of shear rates 0.1~s$^{-1} \leq \dot \gamma \leq 100$~s$^{-1}$ studied.

To further illustrate the difference between hydrophilic and hydrophobic fumed silica suspensions, and to quantify the shear-thickening magnitude, we introduce a dimensionless local parameter $\beta$ defined as $\beta(\sigma) = \dot \gamma ({\rm d}\eta/{\rm d}\sigma) = {\rm d~{\ln}}(\eta)/{\rm d~{\ln}}(\sigma)$, which is directly associated with the slope of the flow curve $\eta(\sigma)$ similarly to previous studies \cite{Brown:2009,Royer:2016}. By definition, $\beta = 0$ for a Newtonian fluid, whereas $\beta<0$ ($\beta>0$ resp.) for a shear-thinning (shear-thickening resp.) behavior. Finally, the parameter $\beta$ saturates at $\beta \geq 1$, when the suspension goes through DST. The evolution of the suspension viscosity and $\beta$ with shear stress is illustrated respectively in Fig.~\ref{fig2}(a) and \ref{fig2}(b), for a suspension of hydrophilic fumed silica particles in PPG at $\phi = 18.9$\%. For increasing imposed shear stress values, $\beta$ smoothly varies from $-0.5$ to 1, which corresponds to a sequence of rheologically-distinct behaviors of the silica suspension from shear thinning (\textcolor{blue}{$\blacktriangledown$}) to Newtonian ($\square$), CST (\textcolor{orange}{$\blacktriangle$}) and finally DST (\textcolor{red}{$\star$}). By repeating such analysis for the flow curves of all measured formulations (i.e., different volume fractions), a rheological state diagram of $\sigma$ vs. $\phi$ for the suspension of hydrophilic fumed silica particles with $R_{\rm u} = 25$ nm is obtained as shown in Fig.~\ref{fig2}(c). Black circles correspond to individual experimental data points, whilst colors encode the local value of $\beta$($\phi$,$\sigma$). Negative values scale from deep blue (indicating pronounced shear thinning, $\beta = -0.5$) to light blue (weakly shear thinning, $\beta \to 0^-$), white values represent the Newtonian regime ($|\beta| < 0.05$), whilst the extent of shear-thickening is represented by a continuous gradient from yellow ($\beta \sim 0.05$) to red ($\beta \to 1$). We define DST to correspond to experimental values of $\beta > 0.9$. This phase diagram for suspensions of hydrophilic fumed silica clearly shows a combination of shear thinning at low shear stress and CST above a critical volume fraction ($\phi_{\rm CST} \sim 5$\%). At larger volume fractions, DST appears for stresses larger than $\sigma_{\rm DST}\simeq 30$~Pa.

The analogous phase diagram of $\sigma$ vs. $\phi$ for suspensions of hydrophobic fumed silica particles is illustrated in Fig.~\ref{fig2}(d). In agreement with Fig.~\ref{fig1}(b), $\beta < 0$ over the entire compositional range studied, which highlights the absence of any CST or DST up to $\phi=20$\%. These results demonstrate the key role of the particle surface chemistry on the macroscopic rheology. Indeed, under strong shear, hydrophilic fumed silica particles brought into close proximity can dynamically form inter-particle hydrogen bonds due to the presence of the hydroxyl groups at the surface of the particles \cite{Raghavan:1997,Raghavan:2000}. Hydrogen bonds have been previously reported to enhance shear thickening in dense suspensions \cite{Comtet:2017,James:2018,James:2019b,Son:2019}. It is clear that coating these hydrophilic particles with alkyl-silane chains results in hydrophobic particles [Fig.~\ref{fig1}(e)], and inhibits the creation of these interparticle hydrogen bonds causing the complete extinction of the shear-thickening transition over this range of volume fraction.

\begin{figure*}[t!]
\centering
\includegraphics[width=0.9\linewidth]{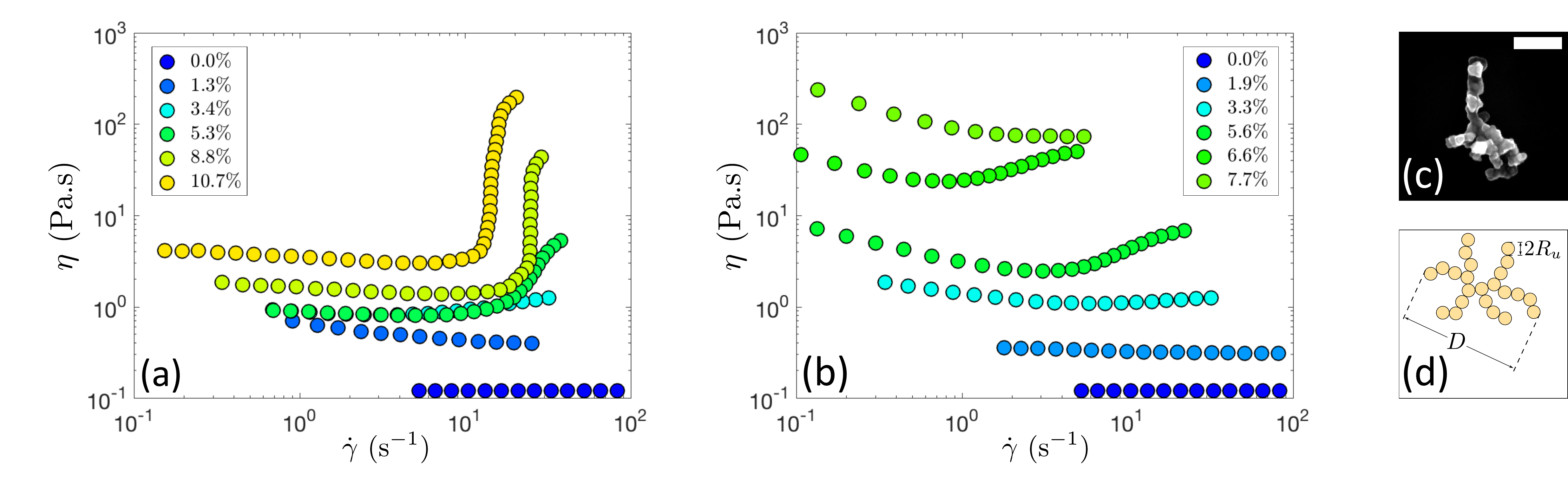}
\caption{(color online) Rheology of suspensions formulated from rougher fumed silica particles. (a) Viscosity $\eta$ of a suspension of hydrophilic fumed silica particles ($R_{\rm u} =10$~nm) vs. the shear stress $\sigma$. The suspension exhibits shear thinning in the dilute regime (blue) before showing CST for $\phi$ = 3.4\% and 5.3\% up to DST at a volume fraction of 8.8\%. (b) Viscosity $\eta$ of a suspension of rougher hydrophobic fumed silica particles ($R_{\rm u} =10$~nm) vs. the shear stress $\sigma$. The suspension exhibits moderate CST and develops a yield stress at high volume fraction. (c) SEM image of a representative hydrophobic particle composed of primary units of size $R_{\rm u} =10$~nm. The scale bar represents 150~nm. (d) Sketch depicting the geometry of these highly-textured fumed silica particles. 
\label{fig3}}
\end{figure*} 

Recently James \textit{et al.} have shown that particle geometry is also important in controlling the range of volume fractions for which shear thickening is observed \cite{James:2019}. To clarify the important contribution of particle surface roughness to the above picture, we also perform a corresponding series of rheological experiments on rougher particles. These fumed silica particles  are composed of a larger number of nodules ($N_{\rm u}\simeq 150$) of smaller size $R_{\rm u}\simeq 10$~nm corresponding to a markedly higher specific surface area (see section~\ref{figS1}). However, these rougher particles possess the same global size $D=300$~nm as the less rough particles (with $R_{\rm u}\simeq 25$~nm) studied in Figures~\ref{fig1} and~\ref{fig2}. As reported in figure~\ref{fig3}(a), suspensions of these rougher hydrophilic particles exhibit flow curves similar to that previously described for the less tortuous (lower specific area) fumed silica [Fig.~\ref{fig1}(c)]. While only shear-thinning behavior is noticed in the dilute regime, CST now appears at $\phi_{\rm CST} \simeq 3.4$\% and DST now occurs for these highly-textured particles at a volume fraction as low as $\phi_{\rm DST} \simeq 8.8$\%. These values are significantly lower than those reported for the less rough particles [Fig.~\ref{fig2}(c)].

These highly-textured fumed silica particles can also be made hydrophobic by silanization [Fig.~\ref{fig3}(c)]. The corresponding rheology, illustrated in Fig.~\ref{fig3}(b) for different volume fractions, shows strong similarity with the case of the less rough hydrophobic particles discussed in Fig.~\ref{fig1}(f). The abrupt discontinuous shear-thickening transition observed for hydrophilic rough particles is spectacularly attenuated by the hydrophobic surface treatment. 
Nonetheless, these suspensions of highly-textured hydrophobic particles still exhibit a weak yet noticeable shear-thickening behavior at large shear rates. However, this CST is only observed over a narrow range of volume fractions $\phi_{\rm CST}\simeq 3.4\% \le \phi \le 7\%$. Beyond this range of volume fractions, the shear thickening is swamped by the rapid rise in the low shear-rate viscosity (denoting the onset of a strong thixotropic behavior and the presence of a yield stress) [see Fig.~\ref{fig3}(b) and the corresponding phase diagram in Fig.~\ref{figS7}]. The existence of a weak but measurable shear thickening effect for these rougher hydrophobic particles at moderate volume fraction demonstrates the possibility of generating some shear thickening even in the absence of attractive chemical forces. This has also been reported for other experimental systems \cite{Osuji:2008}, and recent simulations \cite{Singh:2018,Jamali:2019}. However, local reversible chemical bonds such as interparticle hydrogen bonds dramatically enhance this phenomenon and are essential to induce DST.

From these observations, it is clear that the DST observed in suspensions of the rougher hydrophilic particles reported in Fig.~\ref{fig3}(a) arises from interparticle friction enhanced by reversible chemical interactions, which favor prolonged contact between the textured particles when they are forced into close proximity under strong shear. For the less rough particles, the interparticle friction is too weak to produce shear thickening by itself in the absence of additional short-range chemical interactions as reported in Fig.~\ref{fig1}(b). This influence of the nanometric surface roughness on the rheological behavior of hydrophobic fumed silica suspension allows us to rationalize divergent reports in the existing literature. Indeed, while a weak shear-thickening behavior has been observed with rougher hydrophobic fumed silica particles~\cite{galindo2009shear}, the absence of any shear-thickening regime has been reported in previous studied of less rough hydrophobic particles \cite{raghavan1998composite}. At higher volume fractions, these high surface area particles slowly assemble into a colloidal gel through long range hydrophobic interactions \cite{Israelachvili:1984,Meyer:2006} (see section~\ref{figS5}).

\begin{figure}[t!]
\centering
\includegraphics[width=0.8\linewidth]{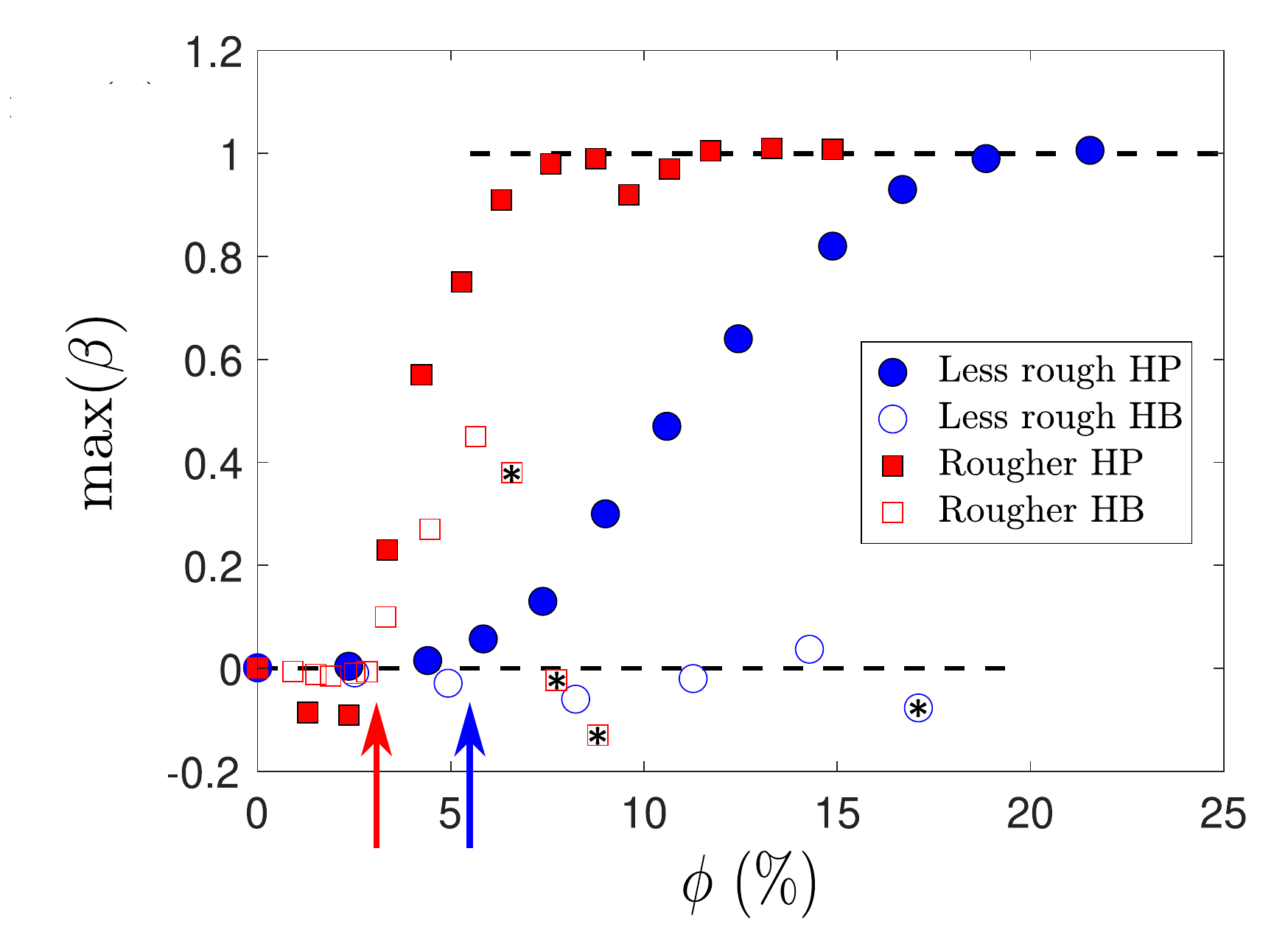}
\caption{(color online) Maximum value of $\beta$ vs. volume fraction $\phi$ determined for four different types of particles characterized by their texture $R_{\rm u}=10$~nm (rougher particles, red squares) or 25~nm (less rough particles, blue circles), and their surface chemistry: hydrophilic (HP, filled symbols) or hydrophobic (HB, open symbols).  Discontinuous shear thickening ($\beta \xrightarrow{} 1$) is observed with hydrophilic particles regardless of the roughness. For the hydrophobic particles, weak continuous shear thickening (CST) is observed for sufficiently rough topography only beyond a critical volume fraction. The continuous shear thickening appears at a critical volume fraction $\phi_{\rm CST}$, whose value depends on the topography of the silica particles: $\phi_{\rm CST} \simeq 3.4$\% (resp. 5\%) for $R_{\rm u} = 10$~nm (resp. 25~nm). The onset of CST is marked by a red (resp. blue) arrow on the graph. Dashed horizontal lines mark the specific values 0 and 1 of ${\rm max}(\beta)$ defining onsets for CST and DST, respectively. The black stars within certain hollow symbols indicate the development of a yield stress at sufficiently high volume fractions.
\label{fig4}}
\end{figure}

In summary, the differences in rheological behavior between the suspensions of particles with two different surface roughness and two different chemical philicities show that the microscopic mechanism for shear thickening is based on a synergy between solid friction and reversible hydrogen bonding; the former being the primary factor to generate at least weak continuous shear thickening and the latter being crucial for developing discontinuous shear thickening at moderate volume fractions. Suspensions formulated from each type of fumed silica particles can also be distinguished by the critical conditions required for the onset of CST and DST. These differences can be easily visualized by plotting the maximum of the slope parameter $\beta$ as a function of the volume fraction $\phi$. The results for the four distinct families of fumed silica suspensions investigated in this article are represented in Fig.~\ref{fig4}.

For suspensions of the rougher silica particles ($R_u$ = 10 nm), CST is observed for volume fractions $\phi \geq \phi_{\rm CST} \simeq 3.4$\%, regardless of whether they have hydrophilic or hydrophobic surface treatments, suggesting this shear-thickening transition is driven by frictional interactions between the highly tortuous particles. By contrast, for the less rough silica particles, weak CST is only observed for hydrophilic particles and for $\phi \geq \phi_{\rm CST} \simeq 5$\%.

\begin{figure*}[t!]
\centering
\includegraphics[width=.8\linewidth]{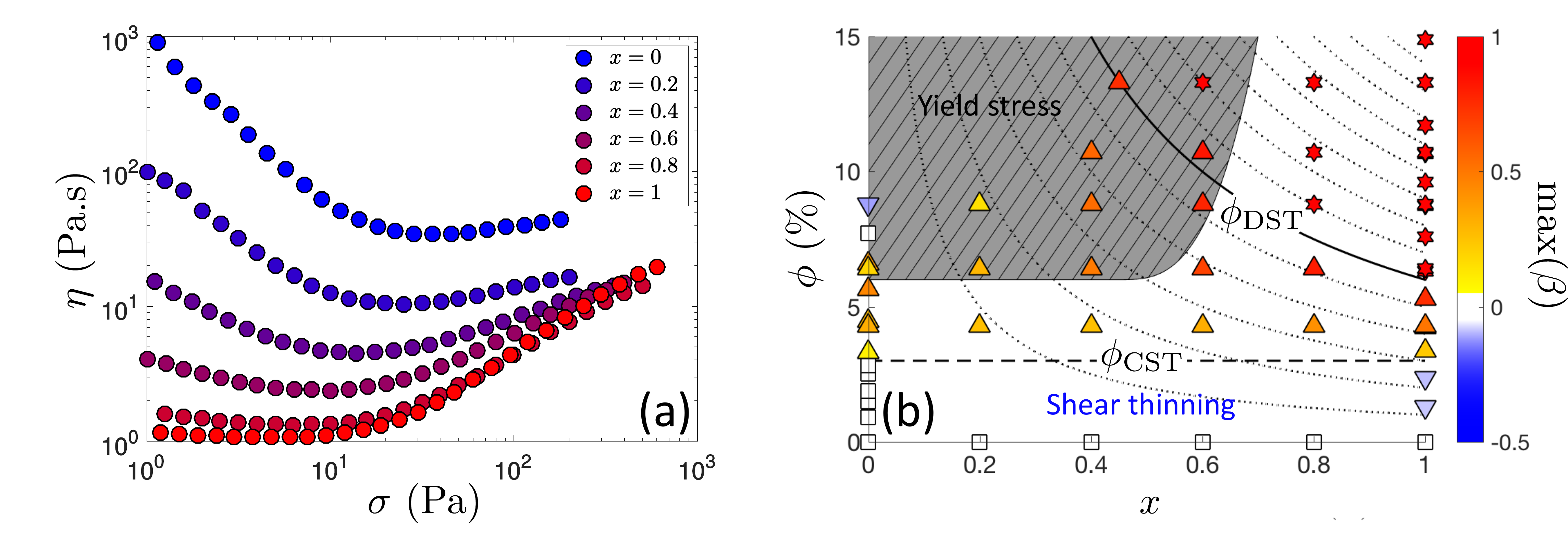}
\caption{(color online) Rheology of mixed suspensions of hydrophilic and hydrophobic silica particles. (a) Flow curves $\eta(\sigma)$ of mixed suspensions of fixed volume fraction $\phi=6.4$\% for different ratio $x$ of hydrophilic to hydrophobic silica particles ($R_{\rm u} = 10$ nm). The rheology shifts from DST for a suspension of pure hydrophilic particles ($x = 1$, \textcolor{red}{$\bullet$}) to a moderate CST for a suspension only composed of hydrophobic particles ($x = 0$, \textcolor{blue}{$\bullet$}). (b) Phase diagram volume fraction $\phi$ vs. hydrophilic ratio $x$ for mixed suspensions of hydrophilic and hydrophobic silica particles. Colors encode the value of ${\rm max}(\beta)$ ranging between $-0.5$ and 1. We observe four different regimes: Newtonian ($\square$), shear thinning (\textcolor{blue}{$\blacktriangledown$}), CST (\textcolor{orange}{$\blacktriangle$} and \textcolor{red}{$\blacktriangle$}) and DST (\textcolor{red}{$\bigstar$}). The top left corner of the diagram that is shaded in grey corresponds to suspensions that show a yield stress. Dotted lines represent iso-contour with a fixed number of hydrophilic particles.  The solid line represents the onset of DST at $\phi^{\rm (HP)}_{\rm DST} = 6\%$.
\label{fig5}}
\end{figure*} 

To understand the origin of the remarkably low values for $\phi_{\rm CST}$ compared to that of other experimental systems reported in the literature, we must consider the hydrodynamic volume influenced by the highly anisotropic and tortuous particles. We define an average inter-particle distance $L$ by allocating a volume $L^3$ to each single particle of solid volume $\Omega_{\rm p}$ such that $\phi \sim \Omega_{\rm p} /  L^3$. The volume $\Omega_{\rm p}$ occupied by one particle is the sum of $N_{\rm u}$ quasi-spherical nodules each of size $R_u$ that have been sintered together, so that $\Omega_{\rm p} \simeq N_{\rm u}(4/3)\pi R_{\rm u}^3$, the interparticle distance $L$ thus scales as follows:
\begin{equation}
L \sim \left[\frac{4\pi N_{\rm u}R_{\rm u}^3}{3\phi} \right]^{1/3} 
\end{equation} 
and decreases with increasing volume fractions of suspended solids. For our textured particles, frictional interaction will become important when the interparticle distance becomes comparable to the particle size, i.e., when $L\simeq D$. This criterion defines the critical volume fraction $\phi_{\rm CST}$ at which shear thickening begins, and depends only on the physical structure of the textured particles (via $D$, $R_{\rm u}$ and $N_{\rm u}$). This estimate of the conditions for the onset of CST is independent of surface chemistry, and thus predicts identical critical conditions for suspensions of rougher particles with both hydrophilic and hydrophobic properties as also noted experimentally in Fig.~\ref{fig4} (see for example \textcolor{red}{$\square$} and \textcolor{red}{$\blacksquare$}). Indeed, both types of particles display identical geometrical characteristics as they only differ by the monolayer silane coating (see Fig.~\ref{figS1}). 

Moreover, this geometric interpretation offers a quantitative estimate of $\phi_{\rm CST}$. We can estimate $N_{\rm u}$ from analysis of multiple SEM pictures (as displayed in Fig.~\ref{figS1}), and find $N_{\rm u} \simeq 150$ and $R_{\rm u} = 10$ nm for the rougher particles, whereas $N_{\rm u} \simeq 25$ and $R_{\rm u} = 25$~nm for the less rough particles. By considering a constant average size of the fumed particles ($D = 300$ nm), in good agreement with our SEM pictures, and using the scaling law described above, we can make an estimate of $\phi_{\rm CST}$ that is in good agreement with the experimental results, i.e., $\phi_{\rm CST}^{\rm th} \simeq 3$\% for $R_{\rm u} = 10$ nm (vs. observed value of $\phi_{\rm CST} \simeq 3.4$\%) and $\phi_{\rm CST}^{\rm th} \simeq 6$\% for $R_{\rm u} = 25$ nm (vs. observed value of $\phi_{\rm CST} \simeq 5$\%). Despite its simplicity (and the uncertainty on the value of $N_{\rm u}$), this geometrical approach provides a simple rationalization and scaling for the surprisingly low critical volume fraction required for obtaining shear thickening with these highly textured particles. Furthermore, the proposed criterion for CST being based on frictional contact between particles~\cite{Wyart:2014,Morris:2018,guazzelli2018rheology,morris2020shear}, our observations and scaling are consistent with a mechanism involving solid friction that is amplified by short-range ($< 1$ nm) reversible physico-chemical interactions~\cite{more2021unifying} (such as hydrogen bonds~\cite{wendler2010estimating}) rather than a scenario based purely on hydrodynamic forces, which might be expected to dominate at long range ($>100$ nm) \cite{Bossis:1989,Kalman:2009,Cheng:2011, morrone2012interplay}. Although moderate CST can be obtained in the absence of hydrogen bond interactions (\textcolor{red}{$\square$} in Fig.~\ref{fig4}), we find that is fully suppressed over the range of volume fractions studied by reducing the nanometric roughness of the particles (\textcolor{blue}{$\circ$} in Fig.~\ref{fig4}), thus providing a key way of tuning the solid friction. Regardless of the scale of the individual particle roughness, the presence of local hydrogen bond interactions is essential to achieve the extreme DST documented in Fig.~\ref{fig1}(c), Fig.~\ref{fig2}(c) and Fig.~\ref{fig3}(a) ($\beta \to 1$).

\textit{Mixtures.-} To offer some insights on the competitive effects of particles with differing surface properties on the shear-thickening transition, we now dilute our suspensions of hydrophilic (HP) fumed silica by substituting some fraction of particles with hydrophobic (HB) fumed silica particles of identical size ($D \approx 300$~nm) and identical topographic features ($R_{u} = 10$~nm). The relative amount of hydrophilic to hydrophobic particles is quantified by the ratio $x = V_{\rm HP}/(V_{\rm HP} + V_{\rm HB})$, where $V_{\rm HP}$ (resp. $V_{\rm HB}$) denotes the volume of hydrophilic (resp. hydrophobic) particles such that $V_{\rm HP} = N_{\rm HP} {\Omega}_{\rm{p}}$ with $N_{\rm HP}$ the number of hydrophilic particles. Figure~\ref{fig5}(a) shows the viscosity of six suspensions of mixed HP and HB silica particles (with $R_{\rm u} = 10$~nm) as a function of the shear stress $\sigma$ at a fixed volume fraction of $\phi= (V_{\rm HP}+V_{\rm HB})/V= 6.4$\%, with $V$ the total volume of the suspension. At $\phi$ = 6.4\%, the suspension of purely hydrophobic particles ($x = 0$) exhibits only a weak continuous shear-thickening behavior  (see Fig.~\ref{fig4}), whereas the suspension of solely hydrophilic particles ($x = 1$) shows DST ($\beta = 1$) with $\eta \propto \sigma$, as already discussed in Figures~\ref{fig3} and~\ref{fig4}. The suspensions of intermediate compositions ($x$ = 0.2, 0.4, 0.6, and 0.8) show a smooth variation of intermediate rheological response between these two limiting cases. In terms of quantifying shear thickening, the slope parameter $\beta$ allows us to determine how the magnitude of shear thickening is shifted from DST to weak CST as we progressively substitute the attractive hydrophilic particles with hydrophobic ones (see Fig.~\ref{figS8}). By exploring systematically the effect of $\phi$ and $x$, we can construct the rheological state diagram reported in Fig.~\ref{fig5}(b), which captures the various rheological regimes of these mixed compositions. For each formulation, the maximum value of the slope parameter, ${\rm max}(\beta)$, is reported as a function of $\phi$ and $x$. Using the same color scheme and symbols as in Fig.~\ref{fig2}, we observe that low total volume fractions always lead to a shear-thinning behavior (\textcolor{blue}{$\blacktriangledown$}), which is more pronounced in the presence of hydrophilic particles. Above a critical volume fraction ($\phi_{\rm CST} \simeq 3.4$\% for these rough particles with $R_{\rm u} = 10$~nm), continuous shear thickening appears (\textcolor{orange}{$\blacktriangle$} and \textcolor{red}{$\blacktriangle$}) for all compositions regardless of $x$; however, the magnitude of the shear thickening response increases with $x$ in agreement with our previous observations. DST (\textcolor{red}{$\star$}) is observed at larger volume fractions ($\phi > 12$\%) in the regime where most of the particles are hydrophilic ($x > 0.5$). Finally, for large volume fractions of hydrophobic particles ($\phi > 6$\% and $x<0.6$), long-range colloidal interactions between the dispersed particles result in a weak gel with a yield stress. This yield stress gives rise to strong shear-thinning in the apparent viscosity that masks any shear-thickening contribution. This dynamic state diagram thus shows the rich variety of rheological regimes achievable by blending different particle chemistries.

As recently described in the case of bi-disperse suspensions, exploring the rheological response of mixtures can also contribute to better understanding of the onset conditions for DST \cite{Madraki:2018}, which we might assert requires a critical number of interparticle frictional contacts. Under this assertion, DST is controlled by exceeding a critical volume fraction of hydrophilic particles $\phi^{\rm (HP)}_{\rm DST} = x \phi_{\rm DST}$ [see thick solid line described by the expression $\phi_{\rm DST} = 6/x$ in Fig.~\ref{fig5}(b), where $\phi^{\rm (HP)}_{\rm DST}\simeq 6$\% for exclusively ($x = 1$) hydrophilic particles with $R_{\rm u} = 10$ nm]. Experiments above this compositional limit indeed show DST for $x \geq 0.8$. However, this is not the case for $x \leq 0.6$ for which we only observe a strong but continuous shear-thickening response at a volume fraction $\phi = 10.7${\%} slightly above the anticipated value of $\phi_{\rm DST} = 6/0.6 = 10${\%}. However, we do ultimately achieve DST at an even higher composition of $\phi = 13.3${\%}. The failure of our simple mixture rule for $x \leq 0.6$ suggest that the incorporation of hydrophobic particles leads to an additional screening of the short-range chemical attractions between hydrophilic particles. This screening effect inhibits the strong sample-spanning frictional interactions required to achieve the DST state but can be overcome by further increasing the volume fraction of hydrophilic particles.

\textit{Conclusion.-} Whilst some degree of shear thickening is often observed in dense suspensions at sufficiently high volume fractions, here we have shown that both CST and DST can be achieved at very low volume fractions by using hydrophilic fumed silica particles possessing high specific surface areas. The surface roughness of these geometrically-irregular particles is the primary control parameter driving the onset of CST, in agreement with the general trends reported in recent experimental and numerical results \cite{Seto:2013,Fernandez:2013,Lin:2015,Royer:2016,Clavaud:2017,Comtet:2017}. However in these extremely rough particles, shear thickening occurs at very low volume fractions and persists over a large range of volume fractions [from 5\% to 23\% in Fig.~\ref{fig2}(c)]. Furthermore we find that the critical shear stress for the onset of shear thickening varies by over an order of magnitude [Fig.~\ref{fig2}(c) and Fig.~\ref{figS10}]. The variation of the critical shear stress with the volume fraction (see section~\ref{figS10}) and the observation of  discontinuous shear thickening over a range of volume fractions far away from the jamming point (see Fig.~\ref{figS11}) offer challenging views to the existing theoretical framework~\cite{Wyart:2014,Morris:2018,guazzelli2018rheology,morris2020shear,rathee2021shear}. These deviations presumably arise from the highly tortuous topography of fumed silica particles and their ability to locally interpenetrate the pervaded volume set by the dimension $D$. This highlights the necessity to extend the constitutive framework developed for regularly-shaped particles to suspensions formulated from more topographically-complex constituents~\cite{guy2020testing}, a question of obvious fundamental and practical interest. Moreover, we also show that DST is only observed in the presence of sufficient levels of hydrogen bonding interactions between the particles. More spectacularly, reducing the level of roughness and eliminating hydrophilic interactions result in the complete suppression of the shear-thickening regime over the range of volume fractions considered here. Finally, we have mapped out a rheological phase diagram for mixtures of hydrophilic and hydrophobic silica particles, which demonstrates the possibility to finely tune the rheological behavior of these fumed silica suspensions by blending hydrophilic and hydrophobic particles. In this way, we screen the strong reversible frictional interparticle interactions and systematically reduce the shear thickening, an observation of obvious practical interest in cement paste rheology \cite{Toussaint:2009}.
This work paves the way for the design of cheaper and more stable shear-thickening fluids requiring lower volume fractions of particles, which is of critical importance for numerous applications such as shock-absorbing composites \cite{Lee:2003}. The optical transparency of these materials (due to the close refractive index match between the fumed silica and the PPG matrix phase) also allows in-situ flow visualization and may provide complementary insights on the suspension microstructure that develops during the discontinuous shear-thickening transition.

\section{ACKNOWLEDGMENTS}

The authors thank R.E. Cohen, F. Galindo-Rosales and S. Manneville for fruitful discussions, as well as the MIT-France and MIT-Portugal seed fund program for financial support. 

\appendix

\renewcommand\thefigure{\thesection S\arabic{figure}}  
\setcounter{figure}{0}  

\section{APPENDIX: SUPPLEMENTAL MATERIALS}

We enclose here supplemental materials following the development of the accompanying paper.

\subsection{S1. Fumed silica particles}

\begin{figure}[t!]
\centering
\includegraphics[width=\linewidth]{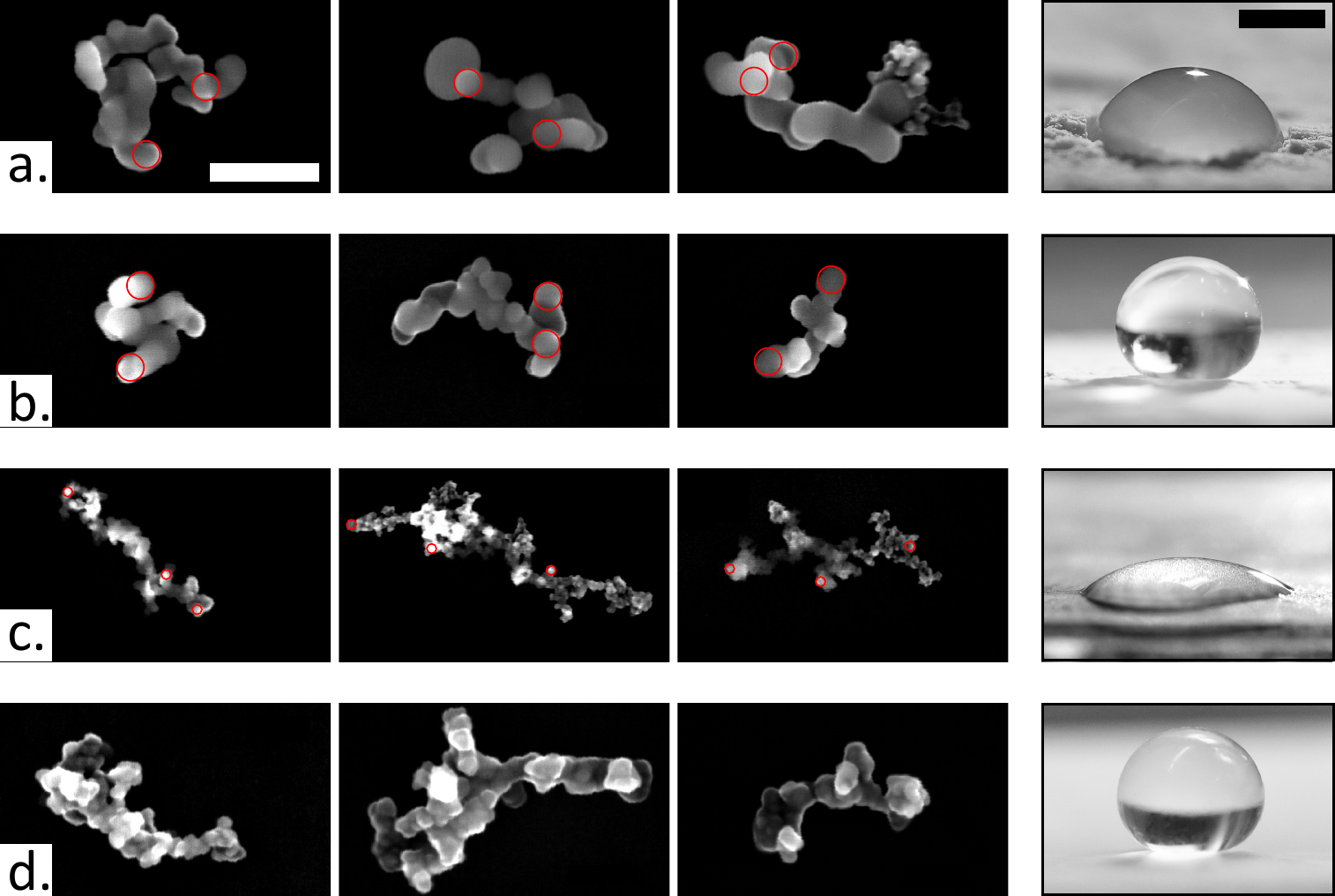}
\caption{Surface properties of fumed silica particles. Each row corresponds to one kind of fumed silica particles and displays SEM pictures of three representative fumed silica particles and an image of a millimetric water drop deposited on a coating obtained with those fumed silica particles. \textbf{(a)} Hydrophilic ``less rough'' fumed silica particles (OX50). \textbf{(b)} Hydrophobic ``less rough'' fumed silica particles (RX50). \textbf{(c)} Hydrophilic ``rougher'' fumed silica particles (AE300). \textbf{(d)} Hydrophobic ``rougher'' fumed silica particles (R812S). The scale bar on SEM picture represents 200 nm. The scale bar on the droplet picture represents 1 mm. The red circles on the SEM images denote the primary particle units or nodules.
\label{figS1}}
\end{figure} 

The fumed silica particles used to prepare the shear-thickening suspensions used in the present study are composed of permanently fused silica nodules (or primary particle units), whose size $R_{\rm u}$ is either 10 nm or 25 nm. SEM images were obtained by mixing a small number of particles ($<$ 10 mg) with 100 mL of isopropanol before depositing a drop of the solution on a piece of flat silicon wafer. Once dried, the particles were imaged with an electronic microscope (FEI Helios NanoLab 600 DualBeam FIB/SEM). Figures~\ref{figS1}(a) and~\ref{figS1}(b) show SEM pictures of, respectively, hydrophilic (OX50, Evonik Industries) and hydrophobic fumed silica particles (RX50, Evonik Industries) of similar geometry, with an average primary unit radius of about $R_{\rm u} = 25$ nm, as better emphasized by the red circles. The RX50 particles are made hydrophobic by silanizing with HDMS (hexamethyldisilazane). Figures~\ref{figS1}(c) and~\ref{figS1}(d) show SEM pictures of the more highly textured (or ``rougher'') fumed silica particles composed of smaller primary units ($R_{\rm u} = 10$ nm) with, respectively, hydrophilic (AE300, Evonik) and hydrophobic (R812S, Evonik) properties. The fractal-like geometries of these different particles explain the differences in the BET surface area (data provided by Evonik): AE300 ($330 \pm 30~\rm{m^2/g}$), OX50 ($50 \pm 15~\rm{m^2/g}$), R812S ($220 \pm 25~\rm{m^2/g}$), RX50 ($35 \pm 10~\rm{m^2/g}$). Indeed, a smaller $R_{\rm u}$ induces a higher specific area $\Sigma _{\rm s}$ scaling as the ratio of the area over the mass $\rho _{\rm s} \Omega _{\rm p}$ of one particle (with $\rho _{\rm s} \approx 2.5~\rm{kg/L}$ the density of silica and $\Omega _{\rm p}$ the volume of an individual particle):
\begin{equation*}
{\Sigma}_{\rm s} \sim \frac{N_{\rm u} 4 \pi {R_{\rm u}}^2}{{\rho _{\rm s}} N_{\rm u} \frac{4}{3} \pi {R_{\rm u}}^3} \sim \frac{3}{\rho _{\rm s} R_{\rm u}}
\end{equation*}
${\Sigma}_{\rm s}$ is ranging from few tens of $\rm{m^2/g}$ to around 100 $\rm{m^2/g}$ for the two primary unit sizes considered in the present work. Importantly, we note that the global size $D$ of the various particles imaged is always of the same order of magnitude, $D \approx 300$ nm.

The surface chemistry of the particle is checked by depositing a layer of particles on the top of double-sided tape. When a water drop is placed on the solid, the coating exhibits a low value of contact angle ($0^{\circ} \leq \theta \leq 50^{\circ}$) for hydrophilic particles, whereas superhydrophobic coatings are achieved ($\theta \geq 150^{\circ}$) with hydrophobic fumed silica due to the combination of hydrophobic chemistry and nanometric roughness.

\subsection{S2. Stability of a fumed silica suspension and repulsive force}

\begin{figure}[b!]
\centering
\includegraphics[width=.8\linewidth]{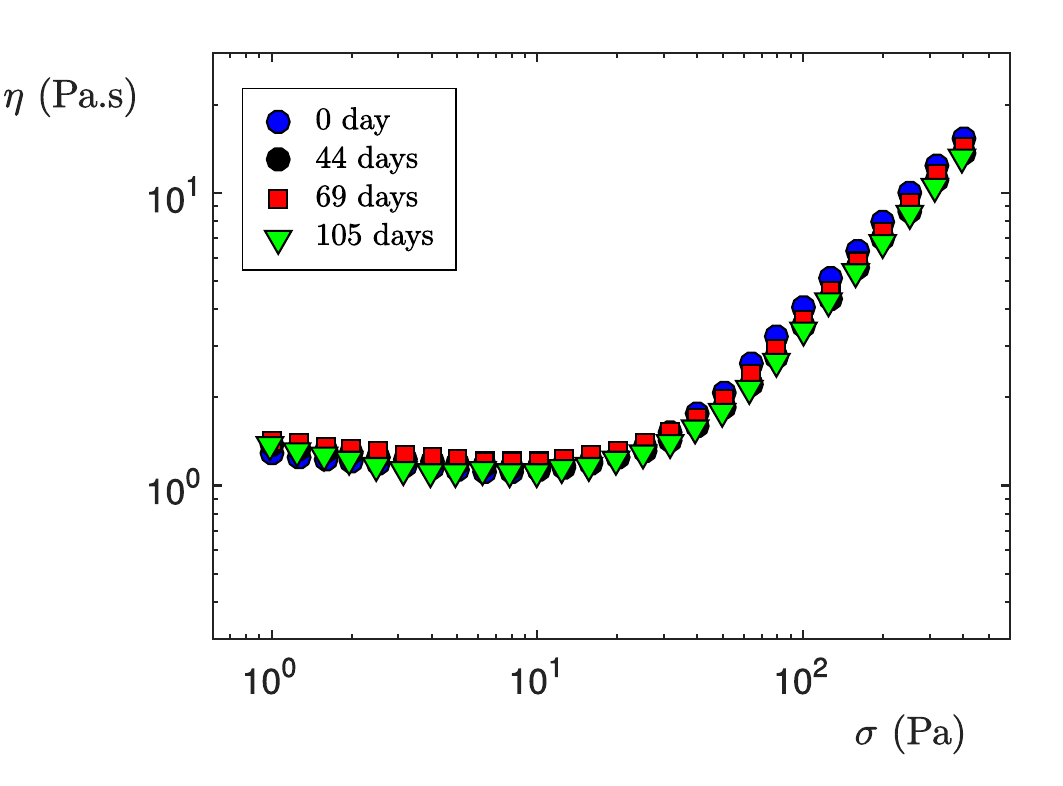}
\caption{Flow curve $\eta ( \sigma )$ of a suspension of ``rougher'' hydrophilic fumed silica particles (AE300, with $\phi = 7.6{\%}$) measured at different times, from 0 to 105 days after sample preparation. Measurements were made using three different cone-plate geometries of diameters 20 mm (red squares after 69 days), 40 mm (0 and 44 days) and 60 mm (green triangles after 105 days).
\label{figS2}}
\end{figure}

Shear-thickening suspensions of fumed silica particles in polypropylene glycol (PPG) are stable over time. The flow curves of a $7.6{\%}$ suspension of rougher hydrophilic fumed silica particles (AE300) determined at different points in time and reported in Fig.~\ref{figS2} illustrate the long-term stability of this suspension of fumed silica particles. Even when measured more than 100 days after preparation, the flow curve $\eta ( \sigma )$ superimposes with the one measured immediately after preparation. The results are also independent of the diameter of the cone-plate test fixture, indicating the lack of any wall slip or other artifacts. Indeed, we measured the same rheological properties with several different rheometric geometries, such as a cone-plate fixture of angle $2^{\circ}$ and diameters 20 mm, 40 mm, or 60 mm.

The long-term stability of fumed silica suspension in polypropylene glycol is without any equivalent in shear-thickening suspensions subject to ageing, drying, migration, or settling. The spectacular stability described in Fig.~\ref{figS2} highlights the presence of a short-range inter-particular repulsive force compatible with the frictional transition scenario~\cite{Wyart:2014}. While the exact nature of the repulsive force remains to be addressed, Brownian motion has been suggested as a possible repulsive force in Brownian suspensions~\cite{mari2015discontinuous, pednekar2017simulation}. Besides, in the case of hydrophilic fumed silica particles in a polar solvent, a short-range non-DLVO repulsive force based on the existence of a solvation layer preventing direct contact between particles has also been suggested~\cite{Raghavan:2000}.

By comparing the viscous stress with the thermal fluctuations, we can estimate a P\'eclet number at the onset of shear-thickening: $\rm{Pe} = {\sigma _{\rm c}} D^3 / {k_{\rm b}} T \sim 200$ with ${\sigma _{\rm c}}\approx 30$ Pa at the onset of shear-thickening in Fig.~\ref{figS2} and $D \approx 300$ nm the global size of a fumed silica particle (see section~\ref{figS1}). The relatively large value of the P\'eclet number fails to predict the transition to shear-thickening (see section~\ref{figS10}). 

\subsection{S3. Determination of the volume fraction $\phi$ by measuring the density $\rho$}

The suspensions of fumed silica particles were prepared by adding a carefully measured mass of dry particles to a controlled mass of polypropylene glycol (PPG 725, Sigma Aldrich). After mixing with a rotor (Heidolph RZR 2102 at 2000 rpm for 5 minutes), air bubbles are removed by placing the samples under vacuum for several hours. The formulation of the suspension is defined by the mass ratio $w$ of fumed silica particles. By carefully measuring the density of such suspensions, it is then possible to determine the true volume fraction ${\phi}$ of the suspension. The density $\rho$ of the suspension corresponds to the ratio of the total mass divided by the volume of the suspension; i.e.  $\rho = ( m_{\rm L} + m_{\rm P} ) / ( {\Omega}_{\rm L} + {\Omega}_{\rm P} )$ with $m_{\rm L}$ and $m_{\rm P}$ (resp. ${\Omega}_{\rm L}$ and ${\Omega}_{\rm P}$) the mass (resp. volume) of the liquid and the particles.

The mass ratio $w$ is defined such that the mass of liquid phase is given by $m_{\rm L} = (1 - w)(m_{\rm L} + m_{\rm P})$. We then obtain a relation between the density of the suspension $\rho$ and the density of the solvent $\rho_{\rm L}$:
\begin{equation*}
\rho = {\rho}_{\rm L} \frac{1 - \phi}{1 - w}
\end{equation*}

Densities were measured precisely using a density meter (Anton Paar DMA 38, accuracy of 0.001 $\rm{kg/L}$) and are reported as a function of the mass ratio $w$ in Fig.~\ref{fig3}(a). By measuring the density $\rho$ and knowing the solvent density $\rho _{\rm L} = 1.007~\rm{kg/L}$, we can precisely evaluate the true volume fraction $\phi$ of any suspension prepared from a known mass ratio $w$:
\begin{equation*}
    \phi = 1 - \frac{\rho}{\rho _{\rm L}}(1-w)
\end{equation*}

\begin{figure}[t!]
\centering
\includegraphics[width=.8\linewidth]{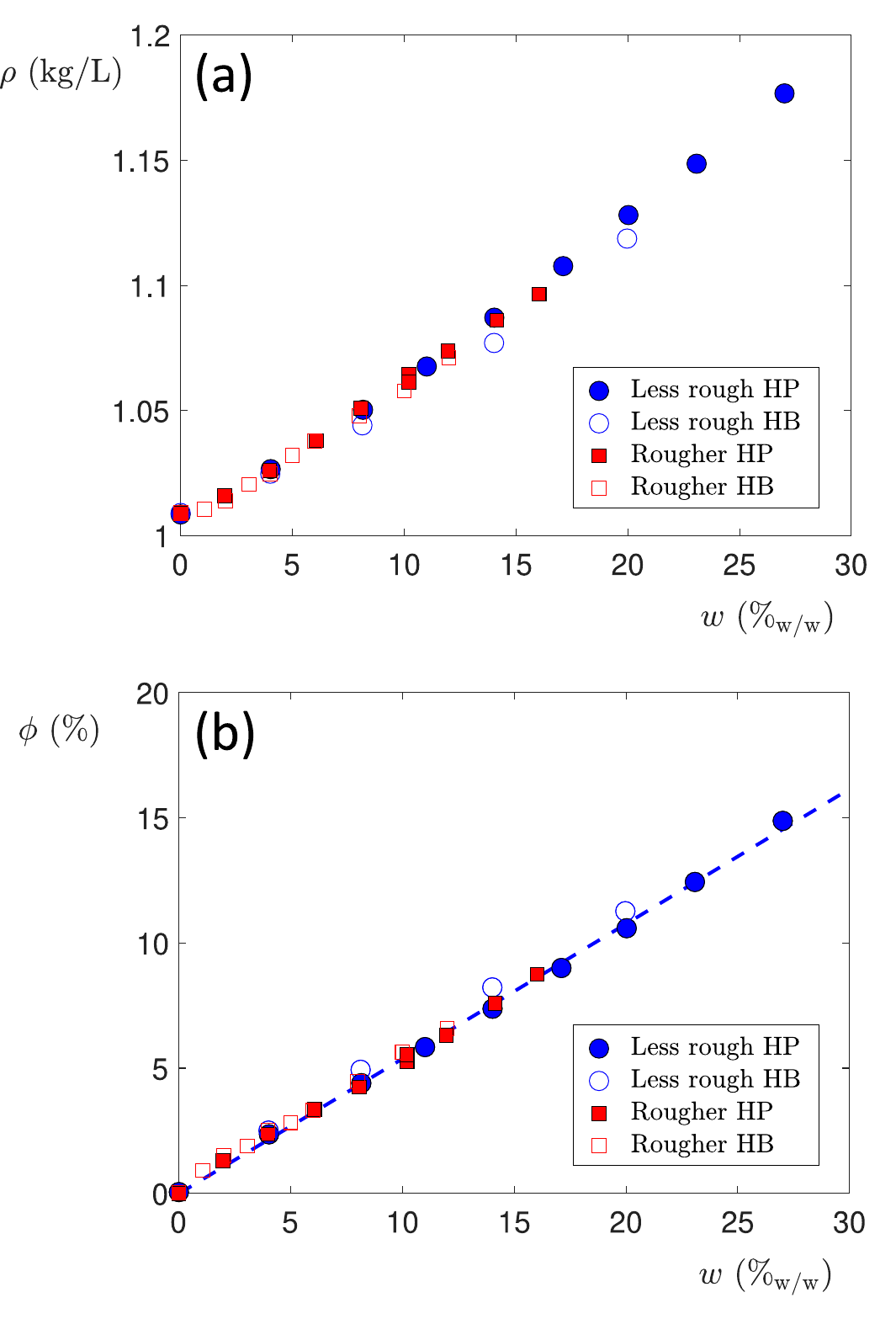}
\caption{Evaluation of volume fraction ${\phi}$. \textbf{(a)} Measured density $\rho$ of the suspension vs. the mass ratio $w$ in particles. \textbf{(b)} Evaluation of the volume fraction $\phi$ of suspensions as a function of the mass ratio $w$. Measurements and estimations were made with both the rougher (red squares, $R_{\rm u} = 10$ nm) and the less rough (blue circles, $R_{\rm u} = 25$ nm) fumed silica particles, including both the hydrophilic (HP, filled symbols) and hydrophobic (HB, open symbols) grades. The blue dashed line in (b) represents the linear fit of the data corresponding to the less rough hydrophilic particles.
\label{figS3}}
\end{figure}

However, at very large volume fractions (e.g., for the ``less rough'' HP particles, $\phi > 20{\%}$), fluid densities are difficult to measure due to the high sample viscosity resulting in entrainment of air bubbles during injection into the densimeter. We then extrapolate densities by considering the linear fit in Fig.~\ref{figS3}(b).

\subsection{S4. Rheological protocol}

\begin{figure}[t!]
\centering
\includegraphics[width=.8\linewidth]{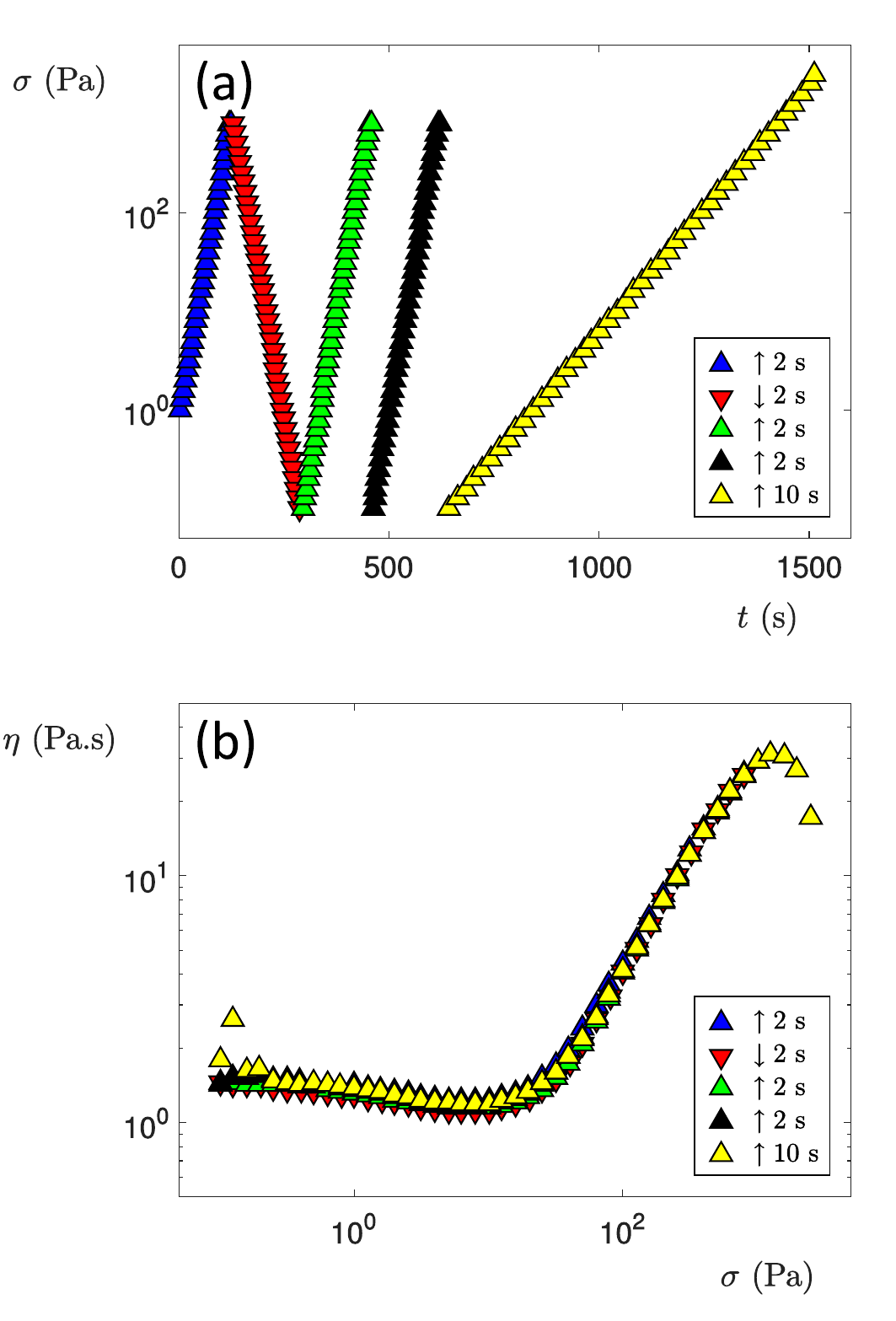}
\caption{Rheological protocol. \textbf{(a)} Stress-controlled experiments are carried out by imposing successive increasing or decreasing ramps in shear stress $\sigma$ with different measurement times. \textbf{(b)} Flow curves $\eta ( \sigma ) $ of a suspension of hydrophilic rougher fumed silica particles in polypropylene glycol ($\phi = 7.6{\%}$) measured from the rheological protocol described in Fig.~\ref{figS4}(a).
\label{figS4}}
\end{figure}

Flow curves were measured using a cone-and-plate geometry ($2^{\circ}$, 40 mm diameter) mounted on a stress-controlled rheometer (DHR3, TA Instruments). We follow the protocol described in Fig.~\ref{figS4}(a), which consists of applying successive up/down ramps in stress. First, an initial ramp is applied by increasing the stress to eliminate the influence of any residual presence of a yield stress. For each measurement point shown, the sample is sheared for 2 s (equilibration time) before averaging the shear rate measurements over the next two 2 s. This first upward stress ramp is followed by a decreasing-stress ramp and two other rising-stress ramps, with each point measured over 2 s. Finally, a last increasing ramp is done with 10 s of equilibration time per point and 10 s of measurement. The five flow curves obtained with such a protocol composed of five successive ramps nicely overlap, as shown in Fig.~\ref{figS4}(b). Flow curves are not affected by the measurement time, which shows the absence of any transient effect in our measurements. Moreover, as the downward ramp in stress overlaps with the others, we do not observe any shear-induced hysteresis or migration in our suspensions, even in the case of DST.

However, in the presence of a yield stress, the viscosity measurements in the lowest range of the shear stress can be affected by the build-up of a solid-like response (see section~\ref{figS5}). Consequently, in the main text, we report flow curves obtained during the second ramp while decreasing the shear stress from the highest value to the lowest value to limit the residual influence of yield stress on the measurement.

\subsection{S5. Yield stress}

\begin{figure}[t!]
\centering
\includegraphics[width=.8\linewidth]{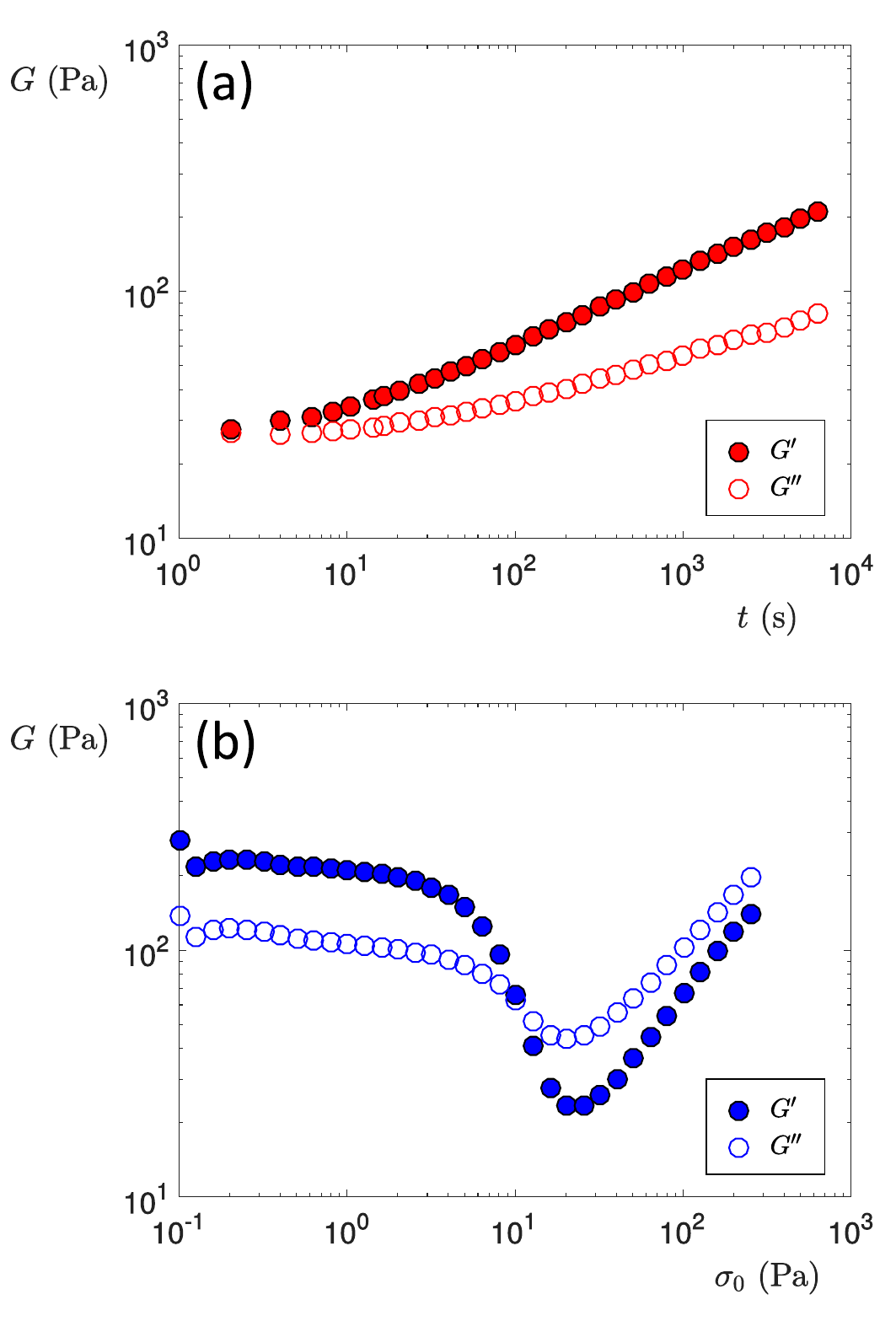}
\caption{Time-dependent behavior in suspensions of hydrophobic rougher fumed silica particles at $\phi = 6.6{\%}$. \textbf{(a)} Rheological aging and evolution of the storage $G^{\prime}$ and loss $G^{\prime \prime}$ moduli vs. time $t$ determined by small amplitude oscillations ($\gamma = 1{\%}$) at a frequency of 1 Hz. \textbf{(b)} Evolution of the storage $G^{\prime}$ and loss $G^{\prime \prime}$ moduli vs. stress amplitude $\sigma _{\rm 0}$ during a stress sweep performed at a frequency of 1 Hz. The cross-over point of $G^{\prime}$ and $G^{\prime \prime}$ at $\sigma _{\rm 0} \approx 10$ Pa defines the static yield stress of the colloidal dispersion. The protocol is applied immediately at the end of the step reported in (a).
\label{figS5}}
\end{figure} 

Some of the suspensions formulated with high volume fractions of the hydrophobic fumed silica exhibit thixotropy and display the growth of a solid-like behavior (through the development of a yield stress) when left at rest (aka ``rheological aging''). Indeed, in the absence of external shear, the yield stress builds-up over timescales of the order of few seconds, as seen in Fig.~\ref{figS5}(a) on a suspension made of hydrophobic rougher fumed silica particles at a volume fraction of $6.6{\%}$. While small amplitude oscillations are imposed to the sample with a strain amplitude $\gamma = 1{\%}$ and a frequency of 1 Hz, we find that the storage modulus, $G^{\prime}$, is larger than the loss modulus, $G^{\prime \prime}$, after 2 s and the system continues to age over the subsequent $10^4$ s. These observations affect the measurement of the flow curves (done with a measurement step time of 2 s), as seen in the lowest range of the shear rate in Fig.~\ref{fig3}(b) in the main text.

To estimate the value of the yield stress, we impose a stress sweep following the aging of the sample over 2 hours [see Fig.~\ref{figS5}(a)]. As seen in Fig.~\ref{figS5}(b), both moduli are constant at small stress values, with $G^{\prime} > G^{\prime \prime}$, indicating the existence of a percolated network i.e., a colloidal gel, before decreasing sharply and crossing at approximately $\sigma _{\rm 0} = 10$ Pa, which corresponds to the static yield stress of the material. The upturn observed at larger stresses is an indication of the subsequent shear-thickening transition observed at large stress amplitudes.

\subsection{S6. Experimental limitations on the determination of flow curves}

\begin{figure}[b!]
\centering
\includegraphics[width=.8\linewidth]{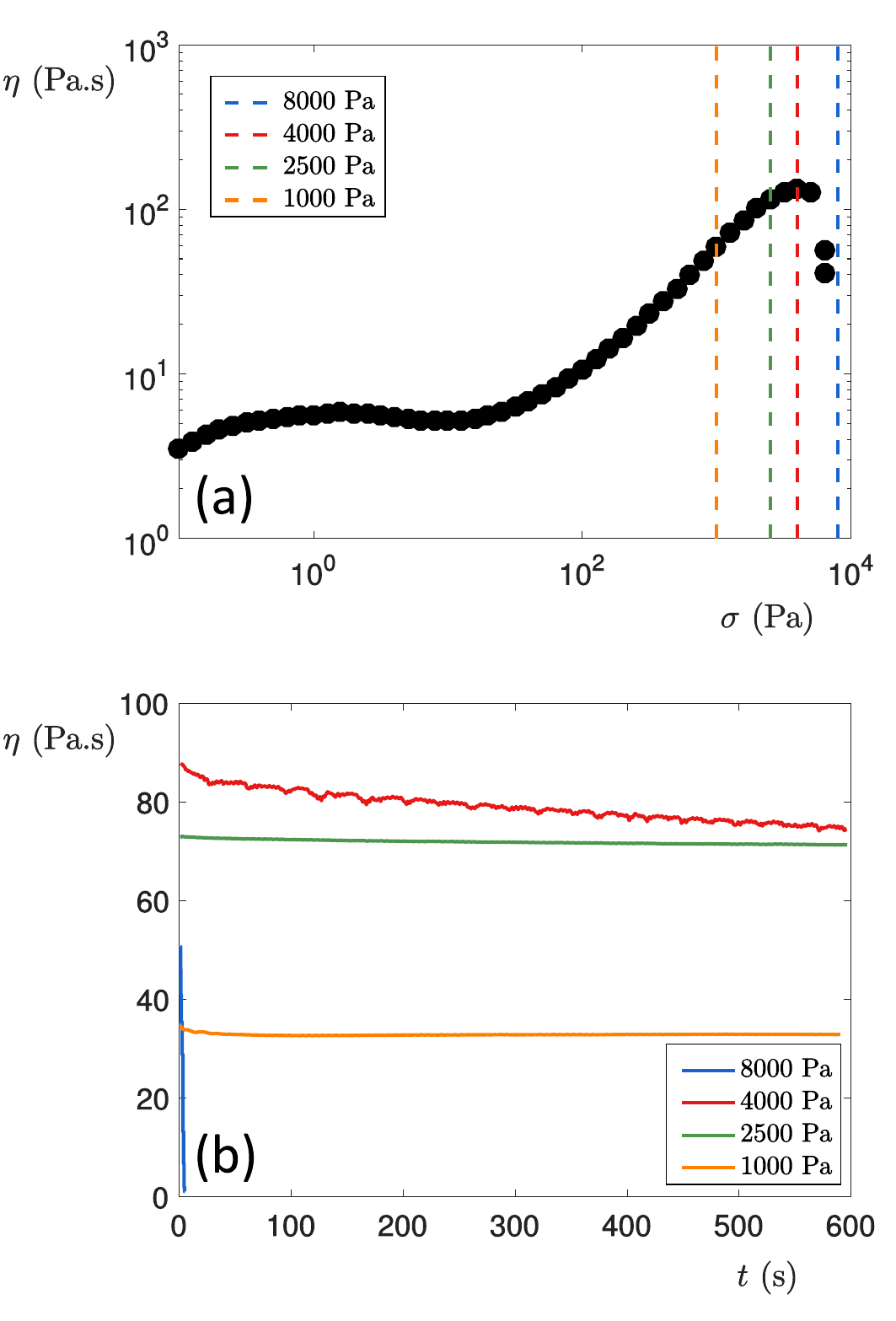}
\caption{Limit of apparent viscosity measurements at high shear stresses illustrated by cone-and-plate meassurements in a suspension of rougher hydrophilic fumed silica particles ($\phi = 9.6{\%}$). \textbf{(a)} Measured viscosity $\eta$ vs. the applied shear stress $\sigma$. The vertical dashed lines mark the stresses explored in (b). \textbf{(b)} Measurement of the apparent viscosity $\eta$ vs. time of shearing $t$ under a range of constant shear stresses (from 1000 to 8000 Pa). 
\label{figS6}}
\end{figure}

The flow curves of a shear-thickening suspension measured with a stress-controlled rheometer usually show a marked decrease of the apparent viscosity reported at very high shear stress. For instance, the viscosity of a suspension of rougher hydrophilic fumed silica particles ($\phi = 9.6{\%}$) exhibits a sudden decrease above a shear stress of 4000 Pa, as seen in Fig.~\ref{figS6}(a).

\begin{figure*}[t!]
\centering
\includegraphics[width=.7\linewidth]{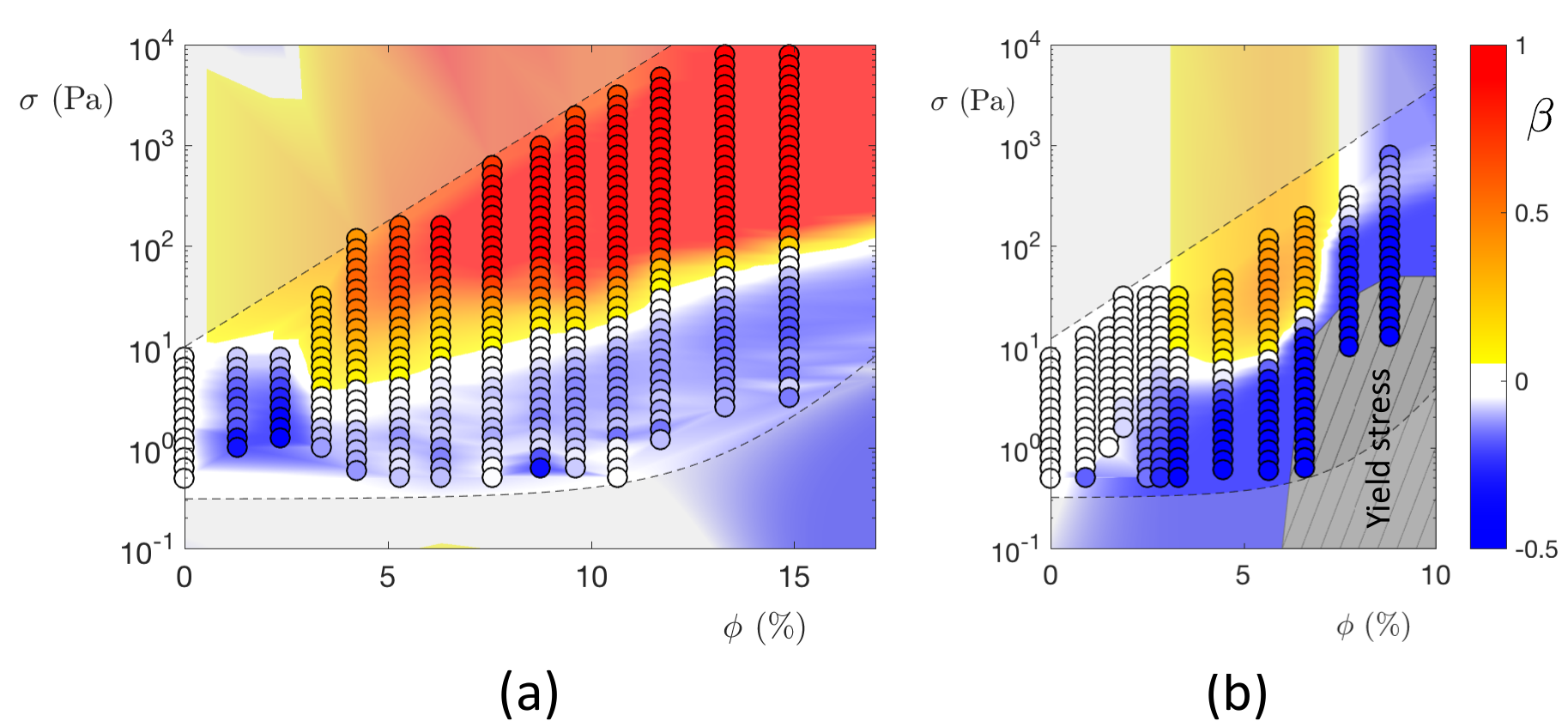}
\caption{Phase diagram showing color contours of the rate of change of viscosity with shear stress $\sigma$ vs. volume fraction $\phi$ of the suspensions of rougher fumed silica particles. \textbf{(a)} Suspensions of rougher hydrophilic fumed silica particles exhibit a multitude of regimes from shear thinning (in blue) to discontinuous shear thickening (in red). \textbf{(b)} Suspensions of rougher hydrophobic fumed silica particles experience a weak continuous shear-thickening regime and the appearance of a yield stress behavior (grey striped area), whereas no discontinuous shear thickening is measured at any applied stress. The dashed lines indicate the measurement limits at low and high shear stress. The colormap encodes the value of the slope parameter $\beta$.
\label{figS7}}
\end{figure*} 

Owing to the remarkable stability of these shear-thickening suspensions, we can measure the evolution in viscosity over timescales of several hundreds of seconds under shearing at constant shear stress, as shown in Fig.~\ref{figS6}(b) (on a linear vertical scale). We observe a stable time-invariant value of the shear-thickened viscosity for shear stresses equal to 1000 Pa and 2500 Pa. However, the apparent viscosity value reported at a shear stress of 4000 Pa decreases by more than 10${\%}$ after only 400 s. More spectacularly, at 8000 Pa, the apparent viscosity drops immediately until reaching a very low value after 4 s, and some material is ejected from the cone-plate geometry. This result indicates that the decrease in apparent viscosity observed at large stresses in the flow curve is not a true material property but an artifact associated with the instability of the underlying rheometric flow. We, therefore, limit the range of shear stresses to $\sigma \leq 2500$ Pa for the flow curve reported in the main text of the manuscript.

\subsection{S7. Phase diagram of suspensions of rougher fumed silica particles}

\begin{figure}[h!]
\centering
\includegraphics[width=.8\linewidth]{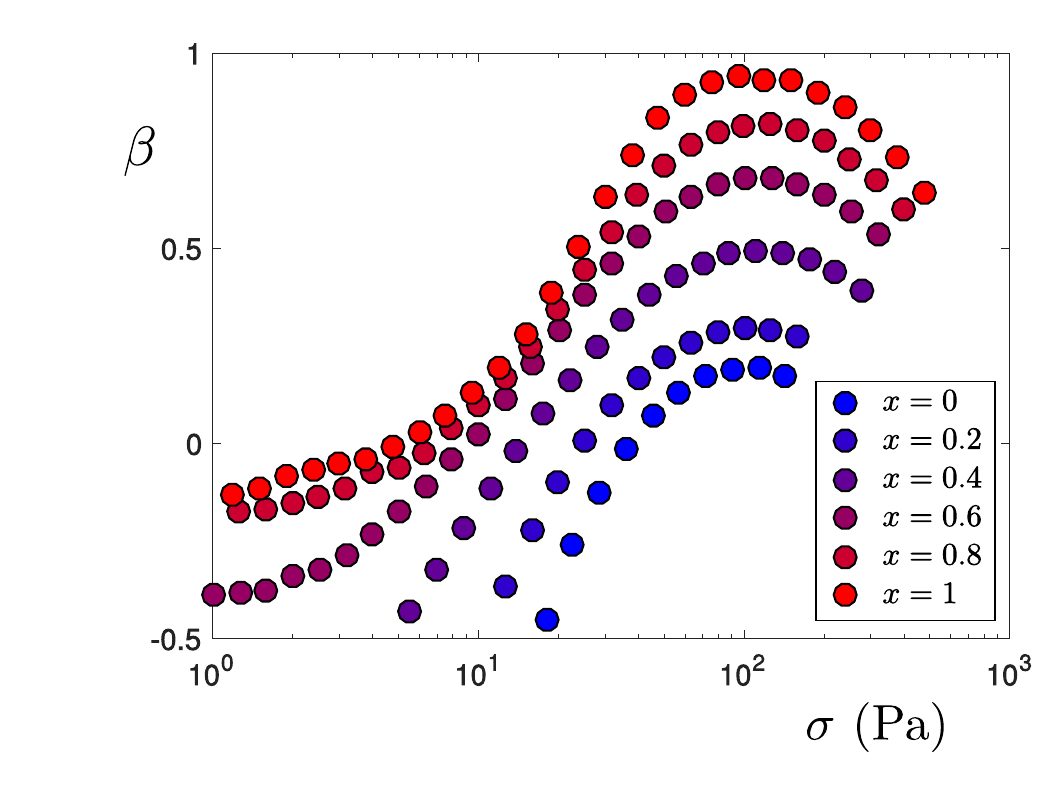}
\caption{Evolution of the slope parameter $\beta$ vs. the imposed shear stress $\sigma$ for various mixtures of rougher fumed silica particles at a fixed volume fraction $\phi = 6.4{\%}$. The mixture ratio $x$ goes from suspensions of purely hydrophobic particles ($x = 0$ in blue) exhibiting weak shear thickening to suspensions of purely hydrophilic particles ($x = 1$ in red) showing discontinuous shear thickening.
\label{figS8}}
\end{figure}

We construct a rheological state diagram (or ``phase diagram'') representing the range of shear-thickening responses observed in rougher ($R_{\rm u} = 10$ nm) colloidal fumed silica suspensions. We perform measurements over a wide range of stresses and volume fractions and extract the local value of the slope parameter $\beta (\sigma , \phi ) = \dot{\gamma} ( \rm{d} \eta / \rm{d} \sigma)$ whose value is encoded in color as shown in Fig.~\ref{figS8}. The phase diagram represents regimes from shear thinning (blue) to Newtonian (white), CST (from yellow to orange) and finally DST (red) for hydrophilic particles [see Fig.~\ref{figS7}(a)]. Black circles represent the data points. The background color was obtained by interpolation. The grey transparent patches at low and high shear stress represent the limit of our measurements whose frontiers are highlighted by dashed lines. For the hydrophobic fumed silica particles [see Fig.~\ref{figS7}(b)], the grey striped area represent the region in which rheological aging is important, as indicated by the appearance of a yield stress (see section~\ref{figS5}).

\subsection{S8. Slope parameter $\beta$ of the flow curves of mixed suspensions}

\subsection{S9. Normal stress}

\begin{figure}[b!]
\centering
\includegraphics[width=.8\linewidth]{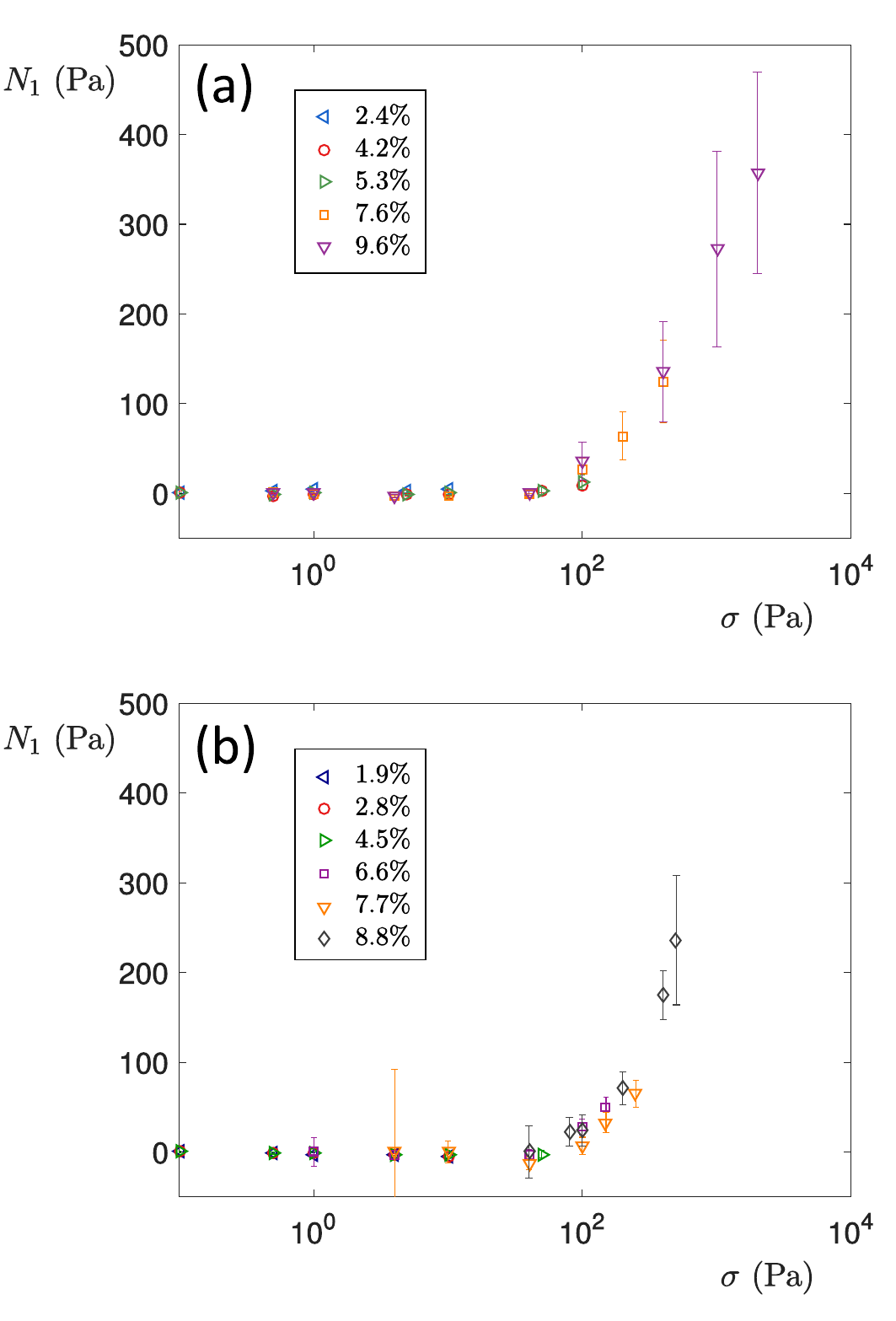}
\caption{The first normal stress difference $N_{\rm 1} ( \sigma )$. \textbf{(a)} Averaged values of $N_{\rm 1}$ (over at least 20 minutes, at constant imposed shear stress) as a function of the shear stress $\sigma$ for different volume fractions $\phi$ ($2.4{\%} \leq \phi \leq 9.6{\%}$) of rougher hydrophilic fumed silica particles. \textbf{(b)} Averaged values of $N_{\rm 1}$ (over at least 20 minutes, at constant imposed shear stress) as a function of the shear stress $\sigma$ for different volume fractions $\phi$ ($1.9 \leq \phi \leq 8.8{\%}$) of rougher hydrophobic fumed silica particles. Error bars represent the standard deviation of the measurement.
\label{figS9}}
\end{figure} 

In addition to measuring the evolution of the viscosity with the shear stress, we also measured the first normal stress difference $N_{\rm 1} ( \sigma )$ by applying a constant shear stress in a cone-and-plate geometry of large diameter (60 mm) for times ranging from several tens of minutes to more than two hours for large stresses and high volume fractions. As noted by Royer \textit{et al.}~\cite{Royer:2016}, the results obtained show strong temporal fluctuations, whose standard deviations are reported in Fig.~\ref{figS9}(a) as error bars in the case of a suspension of rougher hydrophilic fumed silica particles. The stability of the suspension provides a remarkable ability to measure normal forces for durations much larger than other shear-thickening suspensions and thus increases the confidence on the averaged data in Fig.~\ref{figS9}. The mean values of $N_{\rm 1}$ are found to be consistently positive in the shear-thickening regime, in good agreement with frictional mechanisms, as reported by previous studies~\cite{Royer:2016, Lootens:2003, Mari:2014}. Similarly, the weak continuous shear-thickening region reported for the case of hydrophobic rougher fumed silica particles also exhibits smaller but positive values for $N_{\rm 1}$ [see Fig.~\ref{figS9}(b)]. These normal stress signatures are consistent with frictional mechanisms for interparticle interactions.

\subsection{S10. The onset of shear thickening and the frictional transition scenario}

\begin{figure}[b!]
\centering
\includegraphics[width=.8\linewidth]{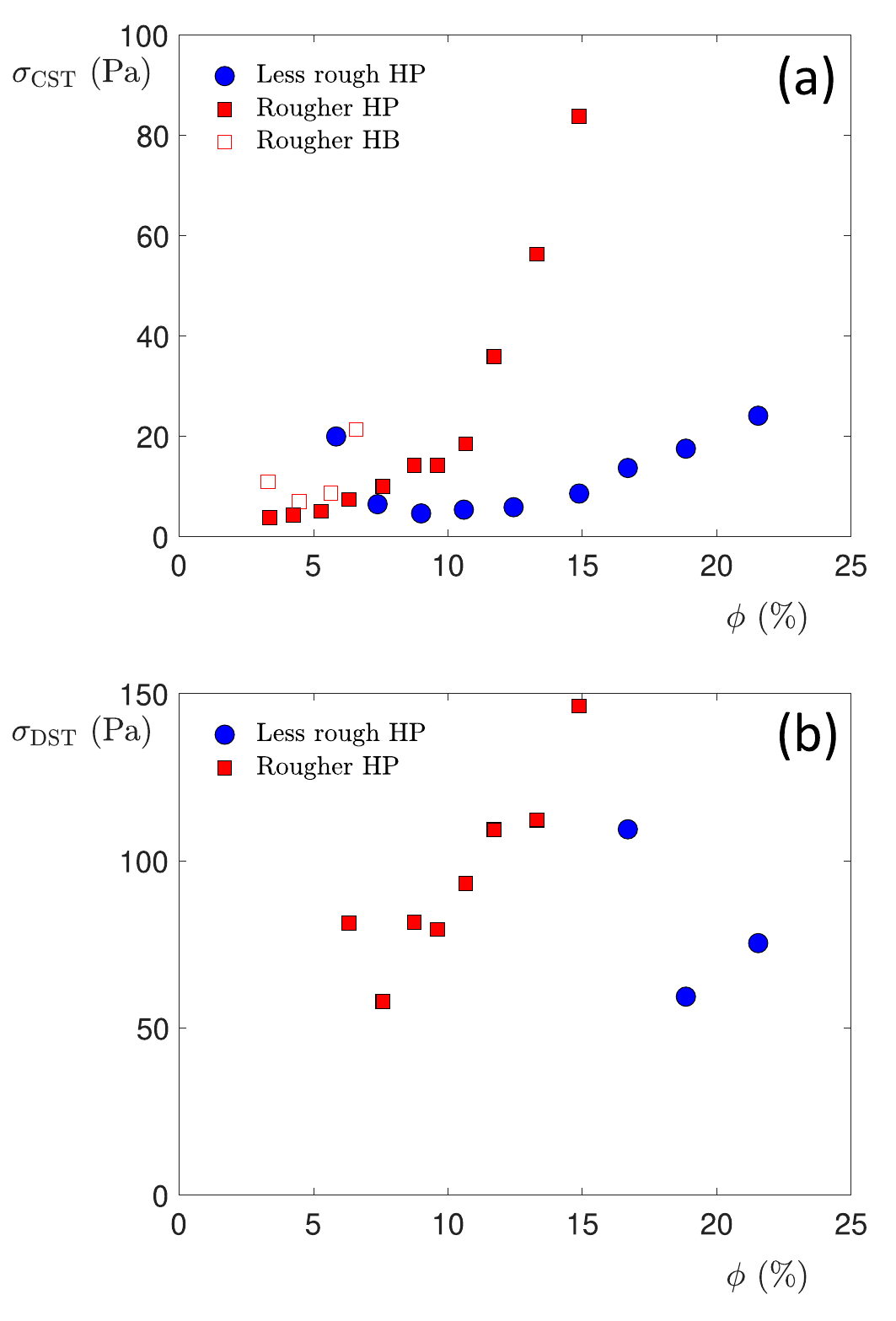}
\caption{Onset of shear-thickening. \textbf{(a)} Critical shear stress ${\sigma}_{\rm{CST}}$ at the onset of Continuous Shear Thickening (CST). CST is observed for both hydrophilic (HP, solid symbols) and hydrophobic (HB, open symbols) rougher (red squares) fumed silica particles and for hydrophilic less rough (blue circles) fumed silica particles. \textbf{(b)} Critical shear stress ${\sigma}_{\rm{DST}}$ at the onset of Discontinuous Shear Thickening (DST). DST is observed only for hydrophilic fumed silica particles either rougher (red squares) or less rough (blue circles).
\label{figS10}}
\end{figure} 

The onset of the shear-thickening has been recently estimated by balancing a repulsive force $F_{\rm{rep}}$ between particles and the shearing force $\sigma D^2$ exerted on a particle of size $D$~\cite{Wyart:2014, clavaud2018shear, morris2020shear}. The repulsion force prevents the contact at low shear-rate, whereas contacts lead to a frictional regime for $\sigma > {\sigma}_{\rm c} = F_{\rm{rep}} / D^2$. This scaling has shown reasonable good agreement with experimental data in various spherical Brownian suspensions~\cite{boersma1990shear, maranzano2001effects} and numerical simulations~\cite{Mari:2014, Seto:2013}. Here, we report the critical shear stresses at the onset of both CST and DST in Fig.~\ref{figS10}. For every formulation, we extract the values of shear stress such as $\beta ( \sigma > {\sigma}_{\rm{CST}} ) > 0.05$ and $\beta ( \sigma > {\sigma}_{\rm{DST}} ) > 0.9$. Despite uncertainties in the lowest range of the volume fractions investigated, all measurements reported in Fig.~\ref{figS10} denote an increase of the critical shear stress with the volume fraction. This result is also apparent in the rheological state diagrams depicted in Fig.~\ref{fig2} in the main text, and section~\ref{figS7} where the lower boundary of both yellow (CST) and red (DST) areas are rising with the volume fraction. In Fig.~\ref{figS10}, the rise of the critical shear stress is particularly apparent in the case of the rougher hydrophilic fumed silica. Both onsets for CST and DST thus show a rapid increase with the volume fraction. This result suggests that the scaling reported in the case of spherical particles do not take into account the specificity of the complex geometry of rough fumed silica particles. In particular, the ability of fumed silica particles to interpenetrate might impose a dependency of the volume fraction on the shearing force.

Besides, one could note the wide range of volume fraction exhibiting CST or DST behavior. This range of volume fraction is without any comparison with other existing systems as seen in Fig.~\ref{figS11}(b).

\subsection{S11. Jamming point and the frictional transition scenario}

\begin{figure}[t!]
\centering
\includegraphics[width=.8\linewidth]{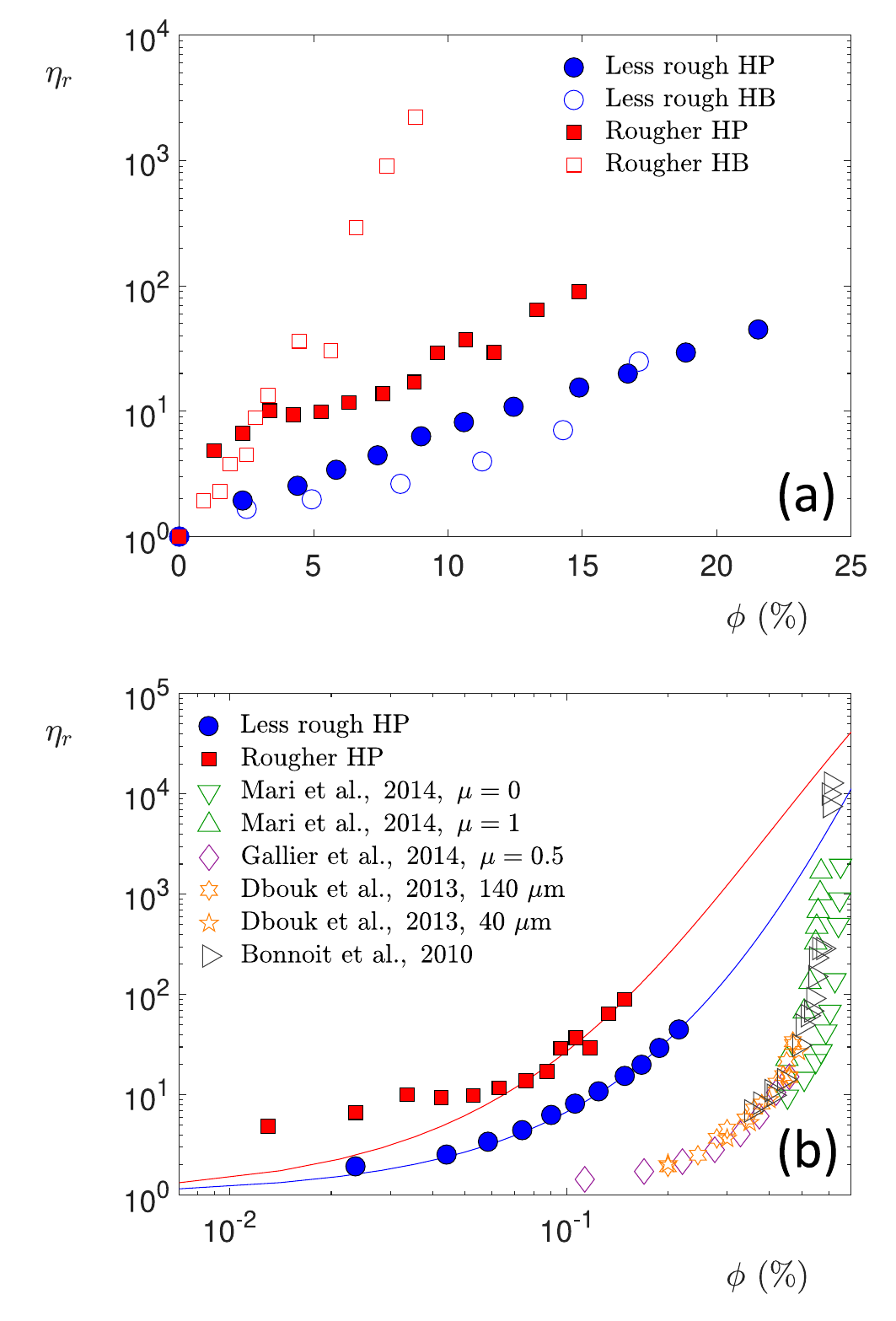}
\caption{Viscosity at rest as function of the particle volume fraction. \textbf{(a)} Relative viscosity at rest $\eta _{\rm r} ( \phi )$ for the four kinds of fumed silica. \textbf{(b)} The relative viscosity at rest can be fitted with the Mooney equation (solid lines) and exhibits a different behavior than the rapid divergence observed for spherical particles (open symbols). Data extracted from~\cite{Mari:2014, guazzelli2018rheology, dbouk2013normal, bonnoit2010inclined, gallier2014rheology}.
\label{figS11}}
\end{figure}

The theoretical framework proposed by Wyart {\&} Cates relates to a transition from separated particles to frictional contacts~\cite{Wyart:2014}. The scenario has been successfully compared with a wide range of shear-thickening responses in non-Brownian and Brownian suspensions, including experimental and numerical results~\cite{Mari:2014,morris2020shear,Clavaud:2017,guazzelli2018rheology, clavaud2018shear}. One possible comparison with the frictional transition scenario arises from the divergence of the viscosity at rest (or minimum viscosity) as a function of the volume fraction $\phi$ at the jamming point $\phi_{\rm J}$. Such divergence has been reported and even compared to an earlier divergence in the presence of friction at lower values of shear rates~\cite{Mari:2014,guazzelli2018rheology, dbouk2013normal, bonnoit2010inclined, gallier2014rheology, lewis1968viscosity}. In Fig.~\ref{figS11}(a), we report the relative viscosity $\eta _{\rm r}$ (at rest) defined as the ratio of the mimimum viscosity $\rm{min} [ \eta ( \phi ) ]$ measured in the range of shear stress investigated over the viscosity of the solvent $\eta ( \phi = 0 )$ for both hydrophilic (solid symbols), and hydrophobic (open symbols) and rougher (red squares) and less rough (blue circles) fumed silica particles. Results for the four fumed silica particles display an increase of the relative viscosity with the volume fraction. However, no clear divergence can be detected in stark contrast with suspensions of spherical particles displayed in Fig.~\ref{figS11}(b). Fitting our data with a power law similar to those reported in the case of spherical particles $\eta _{\rm r} \propto ( \phi _{\rm J} - \phi )^{-2}$~\cite{morris2020shear, guazzelli2018rheology} fails to capture the dependency of $\eta _{\rm r} ( \phi )$. However, our measurements seem to agree with the Mooney equation reported for similar geometries $\eta _{\rm r} = \rm{exp}[a \phi / (1 - b \phi) ]$~\cite{lewis1968viscosity}. A shear-thickening response in a wide range of volume fractions ($\phi \geq 3.3{\%}$ for CST and $6.3{\%} \leq \phi \leq 15{\%}$ for DST with the rougher hydrophilic fumed silica particles) far from the jamming point is in apparent contradiction with Wyart {\&} Cates framework. However, the complex geometry of the fumed silica particle and their ability to interpenetrate remains to be rationalized. While our results suggest that the onset of shear-thickening is set by the interpenetration of adjacent particles (as seen by our scaling and the red and blue arrows in Fig.~\ref{fig4}), a more complex theoretical framework remains to be developed to capture the behavior of fumed silica suspensions.

%

\end{document}